\documentclass[12pt]{article}
\def\be{\begin{equation}}
\def\ee{\end{equation}}
\def\bea{\begin{eqnarray}}
\def\eea{\end{eqnarray}}

\begin{document}

\pagestyle{plain}

\def\e{{\rm e}} \def\cs{\frac{1}{(2\pi\alpha')^2}} \def\CV{{\cal{V}}} \def%
\haf{{\frac{1}{2}}} \def\tr{{\rm Tr}} \def\' \def\p{\partial} \def%
\tphi{\tilde{\phi}} \def\ttheta{\tilde{\theta}} \def\a{\alpha} \def\=\beta 
\def\la{\lambda} \def\barla{\bar{\lambda}} \def\ep{\epsilon} \def\hj{\hat j} 
\def\hn{\hat n} \def\bz{\bar{z}} \def\zk{{\bf{Z}}_k} \def\h1{\hspace{1cm}} 
\def\dd{\Delta_{[N+2k] \times [2k]}} \def%
\ddbar{\bar{\Delta}_{[2k] \times [N+2k]}} \def\u U \def%
\ubar{\bar{U}_{[N] \times [N+2k]}} \def\goes{\rightarrow} \def%
\goal{\alpha'\rightarrow 0} \def\ads\def\ola{\overline {\lambda}} \def%
\oep{\overline {\epsilon}}

\begin{titlepage}
\vskip 2.00cm
\title{
\vbox to\headheight{\hbox to \hsize {\hfill{\small\em\bf BSU-IPP-01/06}}
\vspace{1.5mm}
\footnotesize\hbox to\hsize{\hfill\small\em\ {November~}\number\year~%
}}%
\vspace{2cm}
\bf Energy States of Colored Particle in a Chromomagnetic Field}
\author{Sh. Mamedov\thanks{Email: shahin@theory.ipm.ac.ir \hspace{1mm} \&
\hspace{1mm}  Sh$_-$Mamedov@bsu.az}\\
{\small {\em Institute for Physical Problems, Baku State University, }}\\
{\small {\em Z.Khalilov st. 23, AZ-1148, Baku, Azerbaijan}}}
\date{}

\maketitle

\begin{abstract}
The unitary transformation, which diagonalizes squared Dirac equation 
in a constant chromomagnetic field is found.  Applying this transformation, we find the
eigenfunctions of diagonalized Hamiltonian, that describe the states with definite 
value of energy and call them energy states. It is pointed out that, the energy states
are determined by the color interaction term of the particle with the background 
chromofield and this term is responsible for the splitting of the energy spectrum.
 We construct supercharge operators for the diagonal Hamiltonian, that ensure 
the superpartner property of the energy states. 

\end{abstract}

\vspace{2cm}
PACS: 03.65.-w 

\end{titlepage}

{\Large {\bf Introduction}} \vspace{4mm} 

Classical color field configurations are important for the study of
theoretical problems of non-abelian charged particles. These problems are
connected with the QCD problems [1-3, 11-17] and their study gives a
valuable information on different effects in QCD.

Because color magnetic fields have special significance for the vacuum state
of QCD [2,1], in Refs. [4] and [5] it was considered motion of colored
particle in such fields. For giving a constant and homogenous color
background it was applied the constant vector potentials introduced in [6]
and the squared Dirac equation was solved for this problem. Similarity of
this motion to the motion of an electron in an ordinary magnetic field is
that in both cases we have circular orbits for the motion in the uniform
(chromo)magnetic field and well-known $s$ $p$ $d$ $f$ ... orbitals in a
background field having spherical symmetry. But, the energy spectrum in the
non-abelian problem differs from the one in the abelian case and in
quantized spectrum case it does not look like Landau levels. This is
connected with the matrix structure of the color interaction and color
matrix structure of the Quantum Mechanics of non-abelian charged particles
entirely.

In addition to non-diagonal spin matrices the Dirac equation for these
particles contains the non-diagonal color matrices. These matrices mix
different color and spin components of the wave function in the equations
for these components, which are obtained from the squared Dirac equation. As
a result of such mixing, we cannot write an eigenvalue equation separetaly
for the each color state and for the each spin state in general case of the
background field. Therefore, we are not able to correspond the known wave
functions to the spectrum and to determine thus the energy of each state
with definite value of the spin and color spin. We have not as well a wave
function for description of the states with definite enegy. The way out of
this situation is to diagonalize the squared Dirac operator in the color
spin or combined spin-color spin spaces. For this we should find an unitary
transformation diagonalizing this operator. The wave function of the
particle, as a matrix in these spaces, also will transform under this
transformation. New transformed states will obey the eigenvalue equation for
the diagonalized squared Dirac operator. We aim to find this transformation
and by means of it to construct the states -eigenfunctions of the
diagonalized squared Dirac equation for the considered case of the constant
chromomagnetic background.

Another property of problems in Quantum Mechanics - supersymmetry in Dirac
equation was considered before for this kind of chromomagnetic background
field as well [13, 12, 27]. It is reasonable to reconsider the supersymmetry
for the diagonalized Hamiltonian and to construct the supercharge operators
for the diagonal representation. Knowing the states with definite energy
gives us possibility to study a question of superpartner states in this
supersymmetry and treat thus supersymmetry as a origin of spectrum
degeneration. We suppose, the new supercharge operators will ensure the
superpartneship of the energy states.

\section{Axial chromomagnetic field}

\setcounter{equation}{0}

The Dirac equation for a colored particle in an external color field is
obtained from one for a free particle by the momentum shift:

\begin{equation}
\label{1.1}\left( \gamma ^\mu P_\mu -M\right) \psi =0, 
\end{equation}
where $P_\mu =p_\mu +gA_\mu =p_\mu +gA_\mu ^a\lambda ^a/2$; the $\lambda ^a$
are Gell-Mann matrices describing particle's color spin and within the $%
SU_c(3)$ color symmetry group, $g$ is the color interaction constant and the
color index $a$ runs $a=\overline{1,8}$. Written for the Majorana spinors $%
\phi $ and $\chi $ the equation (1.1) has the well-known form 
\begin{equation}
\label{1.2}\left( \sigma ^iP_i\right) ^2\psi =-\left( \frac{\partial ^2}{%
\partial t^2}+M^2\right) \psi , 
\end{equation}
where the Pauli matrices $\sigma ^i$ describe the particle's spin. Hereafter 
$\psi $ means $\phi $ or $\chi .$ The spinors $\phi $ and $\chi $ have two
components corresponding to the two spin states of a particle $\psi =\left( 
\begin{array}{c}
\psi _{+} \\ 
\psi _{-} 
\end{array}
\right) .$ Each component of $\psi $ transforms under the fundamental
representation of the color group $SU_c(3)$ and has three color components
describing color states of a particle corresponding to the three eigenvalues
of color spin $\lambda ^3$

\begin{equation}
\label{1.3}\psi _{\pm }=\left( 
\begin{array}{c}
\psi _{\pm }(\lambda ^3=+1) \\ 
\psi _{\pm }(\lambda ^3=-1) \\ 
\psi _{\pm }(\lambda ^3=0)\;\; 
\end{array}
\right) =\left( 
\begin{array}{c}
\psi _{\pm }^{(1)} \\ 
\psi _{\pm }^{(2)} \\ 
\psi _{\pm }^{(3)} 
\end{array}
\right) . 
\end{equation}

We are going to continue the study of the motion in the chromomagnetic field
started in [4], where we applied the constant vector potentials introduced
in [6]. Remind here, that for giving an axial chromomagnetic field by the
constant vector potential, components of last one are choosen as below

\begin{equation}
\label{1.4}A_1^a=\sqrt{\tau }\delta _{1a},\ A_2^a=\sqrt{\tau }\delta _{2a},\
A_3^a=0,\ A_0^a=0, 
\end{equation}
where $\tau $ is a constant and $\delta _{\mu a}$ is the Kroneker symbol.

For this $A_\mu ^a$ all components of the field strength tenzor $F_{\mu \nu
}^c=gf^{abc}A_\mu ^aA_\nu ^b$ are equal to zero, except for

\begin{equation}
\label{1.5}F_{12}^3=g\tau =H_z^3, 
\end{equation}
and (1.4) gives a constant chromomagnetic field directed along the third
axes of the ordinary and color spaces. Here $f^{abc}$ are the structure
constants of the $SU_c(3)$ group.

Setting (1.4) and $i\partial \psi /\partial t=E\psi $ in (1.2) we obtain the
following two equations for the $\psi _{\pm }$, which differ from each other
by the sign of the last term in the left hand side: 
\begin{equation}
\label{1.6}\left[ {\bf p}^2+\frac 12g^2\tau I_2+g\tau ^{1/2}\left(
p_1\lambda ^1+p_2\lambda ^2\mp \frac 12g\tau ^{1/2}\lambda ^3\right) \right]
\psi _{\pm }=\left( E^2-M^2\right) \psi _{\pm }. 
\end{equation}
Here $I_2=\left( 
\begin{array}{ccc}
1 & 0 & 0 \\ 
0 & 1 & 0 \\ 
0 & 0 & 0 
\end{array}
\right) $ is the color matrix . Because field (1.5) is directed along the
third axis, $\psi _{\pm }$ describe the spin states with the up and down
projections of $\sigma ^3$ and Hamiltonians defined as $H_{\pm }\psi _{\pm
}=E^2\psi _{\pm }$ correspond to these spin states. Because of non-diagonal $%
\lambda ^1$and $\lambda ^2$ matrices $H_{\pm }$ have not diagonal color
structure and so, we cannot write the eigenvalue equation 
$$
H_{\pm }\psi _{\pm }^{(i)}=E^2\psi _{\pm }^{(i)} 
$$
for pure color states $\psi _{\pm }^{(i)}$ except for the states $\psi _{\pm
}^{(3)}$. The explicit matrix form of the general Hamiltonian in the
combined color and spin spaces is: 
\begin{equation}
\label{1.7}H=\left( 
\begin{array}{cccccc}
{\cal P}^2 & {\cal G}p_{-} & 0 & 0 & 0 & 0 \\ 
{\cal G}p_{+} & {\cal P}^2+{\cal G}^2 & 0 & 0 & 0 & 0 \\ 
0 & 0 & {\cal P}^2 & 0 & 0 & 0 \\ 
0 & 0 & 0 & {\cal P}^2+{\cal G}^2 & {\cal G}p_{-} & 0 \\ 
0 & 0 & 0 & {\cal G}p_{+} & {\cal P}^2 & 0 \\ 
0 & 0 & 0 & 0 & 0 & {\cal P}^2 
\end{array}
\right) . 
\end{equation}
For brevity we have introduced notations ${\cal P}^2${\it $=$}${\bf p}^2+M^2$%
, ${\cal G}=g\tau ^{1/2},$ $p_{\pm }=p_1\pm ip_2.$ Non-diagonality of (1.7)
leads to mixing of the different color states $\psi ^{(i)}$ in the
differential equations obtained from (1.6). But equation for the pure states 
$\psi _{\pm }^{(1),(2)}$ obtained from the (1.7) has the same form for all
the color and spin states [4]. In the cylindrical coordinates the common
equation for $\psi _{\pm }^{(1),(2)}$ has a solution expressed by the Bessel
function $J_m\left( x\right) $: 
\begin{equation}
\label{1.8}\psi _{\pm }^{(i)}\left( {\bf r}\right) =\sum_{m=-\infty
}^{+\infty }\frac 1{2\pi }e^{im\varphi }\exp \left( ip_3z\right) J_m\left(
p_{\bot }r\right) \xi _{\pm }^{(i)}\ . 
\end{equation}
Here $m$ is the chromomagnetic quantum number and $\xi _{\pm }^{(i)}$
includes the spin and color spin parts of wave function. Similarly to the
spin part of wave function, $\xi _{\pm }^{(i)}$ can be choosen as following:%
$$
\xi _{+}^{(i)}=\frac 1{\sqrt{2}}\left( 
\begin{array}{c}
1 \\ 
0 
\end{array}
\right) \zeta ^{(i)},\ \xi _{-}^{(i)}=\frac 1{\sqrt{2}}\left( 
\begin{array}{c}
0 \\ 
1 
\end{array}
\right) \zeta ^{(i)}; 
$$
$$
\ \zeta ^{(1)}=\frac 1{\sqrt{3}}\left( 
\begin{array}{c}
1 \\ 
0 \\ 
0 
\end{array}
\right) ,\ \zeta ^{(2)}=\frac 1{\sqrt{3}}\left( 
\begin{array}{c}
0 \\ 
1 \\ 
0 
\end{array}
\right) ,\ \zeta ^{(3)}=\frac 1{\sqrt{3}}\left( 
\begin{array}{c}
0 \\ 
0 \\ 
1 
\end{array}
\right) . 
$$
Such choosen $\xi $ obeys the normalizing condition below:%
$$
\left| \xi _{+}^{(i)}\right| ^2+\left| \xi _{-}^{(i)}\right| ^2=\left| \zeta
^{(1)}\right| ^2+\left| \zeta ^{(2)}\right| ^2+\left| \zeta ^{(3)}\right|
^2=1. 
$$
Let us remind, that inspite of the same $r-$dependence of all these states,
the different $\psi _{\pm }^{(i)}\left( {\bf r}\right) $ transforms
differently under transformations in the spin and color spaces. This
solution is similar to one for relativistic motion of electron in an axial
magnetic field [7] and in a classical picture gives motion on an circular
orbits [4]. The energy spectrum for (1.6) was found by solving determinant
equation obtained from it and has three continous branches in the case of
infinite motion [10-11]: 
\begin{equation}
\label{1.9}E_{1,2}^2=\left( {\cal P}_{\bot }\mp {\cal G}/2\right)
^2+p_3^2+M^2,\quad E_3^2={\cal P}^2, 
\end{equation}
where ${\cal P}_{\bot }=\sqrt{p_{\bot }^2+{\cal G}^2/4},$ $p_{\bot
}^2=p_1^2+p_2^2.$ Equation for the state $\psi _{\pm }^{(3)}\left( {\bf r}%
\right) $ is the equation for the free particle ${\cal P}^2\psi _{\pm
}^{(3)}\left( {\bf r}\right) =\left( {\bf p}^2+M^2\right) \psi _{\pm
}^{(3)}\left( {\bf r}\right) $ and solution (1.8) can be regarded to it as
well.

For the motion limited by cylinder with the radius $r_0$, the quantized
spectrum has the form [4]: 
$$
\left( E_m^{(N)}\right) _{1,2}^2=\left( \sqrt{\left( \frac{\alpha _m^{(N)}}{%
r_0}\right) ^2+\frac{{\cal G}^2}4}\mp {\cal G}/2\right) ^2+p_3^2+M^2,\quad
\left( E_m^{(N)}\right) _3^2=\left( \frac{\alpha _m^{(N)}}{r_0}\right)
^2+p_3^2+M^2. 
$$
Here $\alpha _m^{(N)}$ are Bessel function's zeros and $N$ is the radial
quantum number. The finite motion solution is connected with one for the
infinite motion by the replacement $p_{\bot }=\alpha _m^{(N)}/r_0$ in (1.8)
and for the wave function we do not sum over the $m$ inasmuch as quantized
energy levels are determined by this quantum number:%
$$
\psi _{\pm }^{(i)}\left( {\bf r}\right) =\frac 1{2\pi }e^{im\varphi }\sin
\left( p_3z\right) J_m\left( \alpha _m^{(N)}r/r_0\right) \xi _{\pm }^{(i)}\
. 
$$
Hereafter, under ${\bf p}$ and under spectrum $E_i^2$ we shall mean their
both continous and quantized values. The spectrum found in such a way, is
not determined by the value of the projection of the color spin operator $%
\lambda ^3$ onto the field (1.5). This means, that states $\psi _{\pm
}^{(1),(2)}$ have not get energy from the branches $E_1$ or $E_2$ definetly.
They can get energy from the both branches of the energy spectrum (1.9), but
with the different probability. It arise question: what are the wave
functions of states having definite energy from the branch $E_1$ or from the
branch $E_2$? It is clear that, for such states we would be able to write
the eigenvalue equation with the spectrum branches (1.9) and this equation
can be written only for the diagonal matrix form of the Hamiltonian. So, we
need the diagonal form of the Hamiltonian (1.7) in the combined spin-color
spin space, in order to write eigenvalue equation by it and then to
determine its eigenfunctions. Since Hamiltonian is the hermitian matrix, it
has diagonal form in the basis of its eigenfunctions and this diagonal form
is unique. We can find the diagonal form of $H^{\prime }$ for the
Hamiltonian (1.7) and then its diagonal elements will correspond to the
spectrum branches from (1.9). By $H^{\prime }$ we will be able to write an
eigenvalue equation with the eigenvalues from (1.9): 
\begin{equation}
\label{1.10}H^{\prime }\Psi ^{\prime }=E_k^2\Psi ^{\prime }. 
\end{equation}
Here $\Psi ^{\prime }$ is the eigenfunction of $H^{\prime }$, which is
different from $\psi _{\pm }^{(1),(2)}.$ In order to get the diagonal $%
H^{\prime }$ from the non-diagonal Hamiltonian $H$ (1.7), we should make
some transformation $U$ in the combined spin-color spin space. The wave
functions $\psi _{\pm }^{(i)}$ will be transformed on this transformation as
well. More precisely, the Hamiltonian (1.7) will get the diagonal form under
some $U$ transformation of the basis vectors of the combined spin-color spin
space. The basis vectors in the combined space are the solutions $\psi _{\pm
}^{(i)}$ (their $\xi _{\pm }^{(i)}$ part), since the eigenfunctions of the
spin and color spin operators $\sigma ^i$ and $\lambda ^a$ are these
functions. Under $U$ transformation of the combined space, the basis
vectors, i.e. $\psi _{\pm }^{(i)}$, transform according to the rule: 
\begin{equation}
\label{1.11}\Psi ^{\prime }=U\Psi ,\qquad \Psi =\left( 
\begin{array}{c}
\psi _{+}^{(1)} \\ 
\psi _{+}^{(2)} \\ 
\psi _{+}^{(3)} \\ 
\psi _{-}^{(1)} \\ 
\psi _{\_}^{(2)} \\ 
\psi _{-}^{(3)} 
\end{array}
\right) . 
\end{equation}
Components of the new basis vector $\Psi ^{\prime }$ will not be the states
with the definite value of projection of spin or color spin, and will be
some superposition of $\psi _{\pm }^{(i)}$. As we stated above, if $%
H^{\prime }$ is diagonal, then basis vectors of this space are the
eigenvectors of this Hamiltonian. So, these components are the states with
the definite value of the energy, because they are eigenfunctions of $%
H^{\prime }$ with the eigenvalue $E_k^2$ from (1.9). We call the components
of $\Psi ^{\prime }$ the $energy\ states$ because of describing the states
having definite energy from the branches $E_k$. Since $H^{\prime }$ is
unique, the transformation matrix $U$ and the basis vector $\Psi ^{\prime }$
are unique as well 
\footnote{We have ignored, the wave functions in Quantum Mechanics are defined up to phase factor. We shall run into this circumstance later.}
.

Under transformation (1.11) the Hamiltonian (1.7), as any matrix in this
space, transforms by similarity transformation: 
\begin{equation}
\label{1.12}U^{-1}HU=H^{\prime }.
\end{equation}
The difficulty of determining of $H^{\prime }$ is that, we have no explicit
form of either $U$ or $H^{\prime }$; two of three matrices in (1.12) are
unknown. But fortunately, it turns out possible to find both of these
matrices, relying on their properties. Firstly, from the hermicity of $%
H^{\prime }$ we can conclude that , the transformation matrix $U$ in
addition to uniqueness, should be unitary: 
$$
H^{\prime \dagger }=U^{\dagger }H^{\dagger }U^{-1\dagger }=H^{\prime
}=U^{-1}HU, 
$$
$$
U^{-1\dagger }=U\Rightarrow UU^{\dagger }=1. 
$$
Equation (1.12) is the basic relation between two Hamiltonian matrices $H$
and $H^{\prime }$. Multiplied from the right hand side by $U$ it has got
more useful form for solving, since contains only the linear relations
between the elements $u_{ij}$ of $U$ matrix: 
\begin{equation}
\label{1.13}HU=UH^{\prime }.
\end{equation}
Another important point for the finding $H^{\prime }$ and $U$ is that, the
unitary transformation does not change determinant and trace of a matrix,
i.e. $H$ and $H^{\prime }$ have the same trace and the same determinant: 
\begin{equation}
\label{1.14}\det H^{\prime }=\det U^{-1}\det H\det U=\det H,\quad \tr H=\tr %
H^{\prime }.
\end{equation}
The matrices $H,$ $H^{\prime }$ and $U$ are matrices of dimensionality $%
6\times 6$. Due to quasidiagonal form of (1.7) we can separately consider
two Hamiltonians $H_{\pm }$ having dimensions $3\times 3,$ instead of one
with dimension $6\times 6$. Of course, equations (1.13) and (1.14) will hold
for both of them, but with different matrices $U_{\pm }$ correspondingly.
The matrices $U_{\pm }$ belong to the color symmetry group $SU_c\left(
3\right) $, so they obey unimodularity condition as well. Solving equations
obtained from (1.13) jointly with the equations of unitarity and
unimodularity we find%
\footnote{Details of this finding is described in Ref. [29] or in the next section for the case considered there. }
the $U_{\pm }$ matrices explicitly: 
\begin{equation}
\label{1.15}U_{\pm }=\left( \frac 1{2{\cal P}_{\bot }}\right) ^{1/2}\left( 
\begin{array}{ccc}
\left( {\cal P}_{\bot }+{\cal G}/2\right) ^{1/2}e^{i\alpha } & \pm
p_{-}\left( {\cal P}_{\bot }+{\cal G}/2\right) ^{-1/2}e^{-i\alpha } & 0 \\ 
\mp p_{+}\left( {\cal P}_{\bot }+{\cal G}/2\right) ^{-1/2}e^{i\alpha } & 
\left( {\cal P}_{\bot }+{\cal G}/2\right) ^{1/2}e^{-i\alpha } & 0 \\ 
0 & 0 & \left( 2{\cal P}_{\bot }\right) ^{1/2}
\end{array}
\right) 
\end{equation}
and establish the diagonal elements of $H_{\pm }^{\prime }$, which
correspondingly equal to 
\begin{equation}
\label{1.16}h_{11}^{\prime }={\cal P}^2-{\cal GP}_{\bot }+{\cal G}^2/2,\ \
h_{22}^{\prime }={\cal P}^2+{\cal GP}_{\bot }+{\cal G}^2/2,\ \
h_{33}^{\prime }={\cal P}^2
\end{equation}
and 
\begin{equation}
\label{1.17}h_{11}^{\prime \prime }={\cal P}^2+{\cal GP}_{\bot }+{\cal G}%
^2/2,\ \ h_{22}^{\prime \prime }={\cal P}^2-{\cal GP}_{\bot }+{\cal G}^2/2,\
\ h_{33}^{\prime \prime }={\cal P}^2.
\end{equation}
Accoding to (1.12) the matrix%
$$
{\cal U}=\left( 
\begin{array}{cc}
U_{+} & 0 \\ 
0 & U_{-}
\end{array}
\right)  
$$
will reduce the Hamiltonian (1.7) to its following diagonal form: 
\begin{equation}
\label{1.18}H^{\prime }=\left( 
\begin{array}{cccccc}
h_{11}^{\prime } & 0 & 0 & 0 & 0 & 0 \\ 
0 & h_{22}^{\prime } & 0 & 0 & 0 & 0 \\ 
0 & 0 & h_{33}^{\prime } & 0 & 0 & 0 \\ 
0 & 0 & 0 & h_{11}^{\prime \prime } & 0 & 0 \\ 
0 & 0 & 0 & 0 & h_{22}^{\prime \prime } & 0 \\ 
0 & 0 & 0 & 0 & 0 & h_{33}^{\prime \prime }
\end{array}
\right) .
\end{equation}
Inspite ${\cal U}$ matrix is the transformation of 6-dimensional spin-color
spin space, indeed it transforms the color space, because it has
quasidiagonal form and does not mix spin indices. Comparing the explicit
forms of the diagonal elements (1.16) and (1.17) with the energy branches
(1.9) it can be easily established the correspondence between them 
\begin{equation}
\label{1.19}
\begin{array}{c}
h_{22}^{\prime },\ h_{11}^{\prime \prime }\longrightarrow E_1^2 \\ 
h_{11}^{\prime },\ h_{22}^{\prime \prime }\longrightarrow E_2^2 \\ 
h_{33}^{\prime },\ h_{33}^{\prime \prime }\longrightarrow E_3^2
\end{array}
.
\end{equation}
According to (1.11) we construct the wave function $\Psi ^{\prime }$, which
will be an eigenfunction of the Hamiltonian (1.18): 
\begin{equation}
\label{1.20}\Psi ^{\prime }={\cal U}\left( 
\begin{array}{c}
\psi _{+}^{(1)} \\ 
\psi _{+}^{(2)} \\ 
\psi _{+}^{(3)} \\ 
\psi _{-}^{(1)} \\ 
\psi _{\_}^{(2)} \\ 
\psi _{-}^{(3)}
\end{array}
\right) =\left( 
\begin{array}{c}
\psi _{+}^{\left( +\right) } \\ 
\psi _{+}^{\left( -\right) } \\ 
\psi _{+}^{(3)} \\ 
\psi _{-}^{\left( +\right) } \\ 
\psi _{-}^{\left( -\right) } \\ 
\psi _{-}^{(3)}
\end{array}
\right) .
\end{equation}
Due to the correspondence (1.19) the desired eigenvalue equation for $%
H^{\prime }$ and $\Psi ^{\prime }$ (1.10) can be written in the following
explicit form : 
$$
\left( 
\begin{array}{cccccc}
h_{11}^{\prime } & 0 & 0 & 0 & 0 & 0 \\ 
0 & h_{22}^{\prime } & 0 & 0 & 0 & 0 \\ 
0 & 0 & h_{33}^{\prime } & 0 & 0 & 0 \\ 
0 & 0 & 0 & h_{11}^{\prime \prime } & 0 & 0 \\ 
0 & 0 & 0 & 0 & h_{22}^{\prime \prime } & 0 \\ 
0 & 0 & 0 & 0 & 0 & h_{33}^{\prime \prime }
\end{array}
\right) \left( 
\begin{array}{c}
\psi _{+}^{\left( +\right) } \\ 
\psi _{+}^{\left( -\right) } \\ 
\psi _{+}^{(3)} \\ 
\psi _{-}^{\left( +\right) } \\ 
\psi _{-}^{\left( -\right) } \\ 
\psi _{-}^{(3)}
\end{array}
\right) =\left( 
\begin{array}{c}
E_2^2\psi _{+}^{\left( +\right) } \\ 
E_1^2\psi _{+}^{\left( -\right) } \\ 
E_3^2\psi _{+}^{(3)} \\ 
E_1^2\psi _{-}^{\left( +\right) } \\ 
E_2^2\psi _{-}^{\left( -\right) } \\ 
E_3^2\psi _{-}^{(3)}
\end{array}
\right) . 
$$
The correspondence between the energy branches $E_k$ and the components $%
\psi _{\pm }^{(\pm )}$, which we have called the energy states, can be found
from the right hand side of the last equality: 
\begin{equation}
\label{1.21}
\begin{array}{c}
E_1^2\longrightarrow \psi _{+}^{\left( -\right) },\ \psi _{-}^{\left(
+\right) } \\ 
E_2^2\longrightarrow \psi _{+}^{\left( +\right) },\ \psi _{-}^{\left(
-\right) } \\ 
E_3^2\longrightarrow \psi _{+}^{(3)},\ \psi _{-}^{(3)}
\end{array}
.
\end{equation}
Having set in (1.20) the explicit form of the matrix ${\cal U}{\bf \ }$ we
find the wave functions of states having energy $E_k$ as a superposition of
spin-color spin states $\psi _{\pm }^{\left( a\right) }$ of non-transformed
space:

$$
\psi _{\pm }^{\left( +\right) }=u_{11}\psi _{\pm }^{(1)}+u_{12}\psi _{\pm
}^{(2)}=\left( 2\widehat{\cal P}_{\bot }\right) ^{-1/2}\left( \widehat{\cal P%
}_{\bot }+{\cal G}/2\right) ^{1/2}e^{i\alpha }\psi _{\pm }^{(1)} 
$$
$$
\pm \widehat{p}_{-}\left( 2\widehat{\cal P}_{\bot }\right) ^{-1/2}\left( 
\widehat{\cal P}_{\bot }+{\cal G}/2\right) ^{-1/2}e^{-i\alpha }\psi _{\pm
}^{(2)}, 
$$
$$
\psi _{+}^{\left( -\right) }=\psi _{-}^{\left( -\right) }=u_{21}\psi _{\pm
}^{(1)}+u_{22}\psi _{\pm }^{(2)}=\mp \widehat{p}_{+}\left( 2\widehat{\cal P}%
_{\bot }\right) ^{-1/2}\left( \widehat{\cal P}_{\bot }+{\cal G}/2\right)
^{-1/2}e^{i\alpha }\psi _{\pm }^{(1)} 
$$
$$
+\left( 2\widehat{\cal P}_{\bot }\right) ^{-1/2}\left( \widehat{\cal P}%
_{\bot }+{\cal G}/2\right) ^{1/2}e^{-i\alpha }\psi _{\pm }^{(2)}. 
$$
Having written the momentum operators $\widehat{p}_i$ in the polar
coordinates $r$ and $\varphi $%
$$
\widehat{p}_1\pm i\widehat{p}_2=-ie^{\pm i\varphi }\left( \frac \partial
{\partial r}\pm \frac ir\frac \partial {\partial \varphi }\right) 
$$
and applying the reccurrent formula for differentiation of the Bessel
functions $J_m\left( x\right) $

$$
J_m^{\prime }\left( x\right) =\frac mxJ_m\left( x\right) -J_{m+1}\left(
x\right) =J_{m-1}\left( x\right) -\frac mxJ_m\left( x\right) 
$$
we easily establish action of these operators on Bessel functions:

\begin{equation}
\label{1.22}\left( \widehat{p}_1\pm i\widehat{p}_2\right) J_m\left( p_{\bot
}r\right) e^{im\varphi }=\pm ip_{\bot }J_{m\pm 1}\left( p_{\bot }r\right)
e^{i\left( m\pm 1\right) \varphi }, 
\end{equation}
$$
\widehat{p}_{\bot }^2J_m\left( p_{\bot }r\right) e^{im\varphi }=\left( 
\widehat{p}_1\pm i\widehat{p}_2\right) \left( \widehat{p}_1\mp i\widehat{p}%
_2\right) J_m\left( p_{\bot }r\right) e^{im\varphi }=p_{\bot }^2J_m\left(
p_{\bot }r\right) e^{im\varphi }, 
$$

$$
\widehat{\cal P}_{\bot }\psi _{\pm }^{(i)}=\sqrt{\widehat{p}_{\bot
}^2+g^2\tau /4}\psi _{\pm }^{(i)}={\cal P}_{\bot }\psi _{\pm }^{(i)}. 
$$
As is seen, action of operators $\widehat{p}_1\pm i\widehat{p}_2$ shift the
state with $m$ to the state with $m\pm 1$. This means that, in the quantized
spectrum case, when the transverse momentum $p_{\bot }$ gets values
determined by the chromomagnetic quantum number $m:$ $p_{\bot }=\alpha
_m^{(N)}/r_0$ and we do not sum over this quantum number, the energy states
will be superposition of the states with the different values of $m$: 
$$
\psi _{\pm }^{\left( +\right) }=\frac{\sin \left( p_3z\right) }{\sqrt{2{\cal %
P}_{\bot }}}\frac{e^{im\varphi }}{\sqrt{2\pi }}\left[ \left( {\cal P}_{\bot
}+{\cal G}/2\right) ^{1/2}e^{i\alpha }J_m\left( p_{\bot }r\right) \zeta
_{\pm }^{(1)}\right. 
$$
$$
\left. \mp ip_{\bot }\left( {\cal P}_{\bot }+{\cal G}/2\right)
^{-1/2}e^{-i\alpha }J_{m-1}\left( p_{\bot }r\right) e^{-i\varphi }\zeta
_{\pm }^{(2)}\right] , 
$$
$$
\psi _{\pm }^{\left( -\right) }=\frac{\sin \left( p_3z\right) }{\sqrt{2{\cal %
P}_{\bot }}}\frac{e^{im\varphi }}{\sqrt{2\pi }}\left[ \mp ip_{\bot }\left( 
{\cal P}_{\bot }+{\cal G}/2\right) ^{-1/2}e^{i\alpha }J_{m+1}\left( p_{\bot
}r\right) e^{i\varphi }\zeta _{\pm }^{(1)}\right. 
$$
$$
\left. +\left( {\cal P}_{\bot }+{\cal G}/2\right) ^{1/2}e^{-i\alpha
}J_m\left( p_{\bot }r\right) \zeta _{\pm }^{(2)}\right] . 
$$
In the continous spectrum case, the transverse momentum $p_{\bot }$ and the
spectrum does not depend on $m$ and we can sum over this quantum number in
energy states, as we made in (1.8). This just leads to replacement of the
momentum operators by their eigenvalues and then the energy states have got
the form:

$$
\psi _{\pm }^{\left( +\right) }=\frac 1{\sqrt{2{\cal P}_{\bot }}}\left( 
{\cal P}_{\bot }+{\cal G}/2\right) ^{1/2}e^{i\alpha }\psi _{\pm }^{(1)}\mp 
\frac{ip_{\bot }}{\sqrt{2{\cal P}_{\bot }}}\left( {\cal P}_{\bot }+{\cal G}%
/2\right) ^{-1/2}e^{-i\alpha }\psi _{\pm }^{(2)}, 
$$
\begin{equation}
\label{1.23}\psi _{\pm }^{\left( -\right) }=\mp \frac{ip_{\bot }}{\sqrt{2%
{\cal P}_{\bot }}}\left( {\cal P}_{\bot }+{\cal G}/2\right)
^{-1/2}e^{i\alpha }\psi _{\pm }^{(1)}+\frac 1{\sqrt{2{\cal P}_{\bot }}%
}\left( {\cal P}_{\bot }+{\cal G}/2\right) ^{1/2}e^{-i\alpha }\psi _{\pm
}^{(2)}.
\end{equation}
It can be found from (1.23) that, all energy states have wave functions with
same module: 
$$
\left| \psi _{\pm }^{\left( \pm \right) }\right| ^2=\frac{{\cal P}_{\bot
}\pm {\cal G}/2}{2{\cal P}_{\bot }}\left| \psi _{\pm }^{\left( 1\right)
}\right| ^2+\frac{{\cal P}_{\bot }\mp {\cal G}/2}{2{\cal P}_{\bot }}\left|
\psi _{\pm }^{\left( 2\right) }\right| ^2 
$$
\begin{equation}
\label{1.24}=\frac 16\left[ \sum_{m=-\infty }^{+\infty }\frac 1{2\pi
}e^{im\varphi }\exp \left( ip_3z\right) J_m\left( p_{\bot }r\right) \right]
^2=F^2(r).
\end{equation}
This means, that a distribution of particles on the states with energy $E_2$
or $E_1$ are same and does not depend on spin or color quantum numbers. This
is a result of invariance of distribution with respect to transformation
(1.11):%
$$
\left| \psi _{\pm }^{\left( +\right) }\right| ^2+\left| \psi _{\pm }^{\left(
-\right) }\right| ^2=\left| \psi _{\pm }^{\left( 1\right) }\right| ^2+\left|
\psi _{\pm }^{\left( 2\right) }\right| ^2. 
$$
As is known, eigenfunctions of any conserved quantity are the eigenfunctions
of Hamiltonian as well. At now we wish to clear what is the conserved
operator in 6-dimensional combined spin-color spin space, eigenfunctions of
which are the found energy states $\psi _{\pm }^{\left( \pm \right) }$? Of
course, sought operator determines the branches of spectrum and will commute
with $H^{\prime }$. In order to find this operator, let us divide
Hamiltonian (1.7) into diagonal part ${\bf p}^2+\frac 12g^2\tau I_2+M^2$,
which does not changes under transformation (1.12), and non-diagonal part
[13], [25] 
\begin{equation}
\label{1.25}\left( \sigma ^iP_i\right) ^2-\left( {\bf p}^2+\frac 12g^2\tau
I_2\right) =g\tau ^{1/2}\left( p_1\lambda ^1+p_2\lambda ^2-\frac 12g\tau
^{1/2}\sigma ^3\lambda ^3\right) ,
\end{equation}
which becomes diagonal under this transformation: 
$$
{\cal U}^{-1}g\tau ^{1/2}\left( p_1\lambda ^1+p_2\lambda ^2-\frac 12g\tau
^{1/2}\sigma ^3\lambda ^3\right) {\cal U}= 
$$
\begin{equation}
\label{1.26}\left( 
\begin{array}{cc}
U_{+}^{-1} & 0 \\ 
0 & U_{-}^{-1}
\end{array}
\right) \left( 
\begin{array}{cc}
\left( \lambda ^bI^b\right) _{+} & 0 \\ 
0 & \left( \lambda ^bI^b\right) _{-}
\end{array}
\right) \left( 
\begin{array}{cc}
U_{+} & 0 \\ 
0 & U_{-}
\end{array}
\right) =\left( 
\begin{array}{cc}
\left( \lambda ^bI^b\right) _{+}^r & 0 \\ 
0 & \left( \lambda ^bI^b\right) _{-}^r
\end{array}
\right) .
\end{equation}
This non-diagonal term is the interaction term of the particle with the
chromomagnetic background. Here the two operators $\left( \lambda
^bI^b\right) _{\pm }=g\tau ^{1/2}\left( p_1\lambda ^1+p_2\lambda ^2\mp \frac
12g\tau ^{1/2}\lambda ^3\right) $ correspond to the two different spin
states of the particle and in the transformed spin-color spin space these
operators has got a diagonal form: 
$$
\left( \lambda ^bI^b\right) _{+}^r=\left( 
\begin{array}{ccc}
-{\cal GP}_{\bot } & 0 & 0 \\ 
0 & {\cal GP}_{\bot } & 0 \\ 
0 & 0 & 0
\end{array}
\right) ,\qquad \left( \lambda ^bI^b\right) _{-}^r=\left( 
\begin{array}{ccc}
{\cal GP}_{\bot } & 0 & 0 \\ 
0 & -{\cal GP}_{\bot } & 0 \\ 
0 & 0 & 0
\end{array}
\right) . 
$$
As is seen from (1.9) and (1.26), the term with $\pm $ sign in the energy
spectrum was appeared by this term, which can be written as the scalar
product of two color vectors 
\footnote{Under vector here we mean eight component quantity in the color space.}
$\lambda ^b$ and $I^b$: 
\begin{equation}
\label{1.27}\left( \lambda ^bI^b\right) =g\lambda ^b\left( A_j^bp_j-\frac
g4\epsilon _{ij3}f^{acb}A_i^aA_j^c\sigma ^3\right) =2g\left( A_jp_j-\frac
12F_{12}\sigma ^3\right) .
\end{equation}
Here $F_{12}=F_{12}^3\lambda ^3/2$. This product is the projection of the
color spin operator $\lambda ^b$ onto the color vector $I^b$ and in the
rotated color space this projection has got the diagonal form with the three
eigenvalues $\pm {\cal GP}_{\bot }$, $0$. Three different branches of energy
spectrum correspond to these three values of $\left( \lambda ^bI^b\right) $
projection. Since operator $\sigma ^3/2$ commute with the general
Hamiltonian (1.7) and consequently, is the conserved quantity, this operator
describes the spin of the particle in the field (1.4). Two Hamiltonians $%
H_{\pm }$ correspond to the two eigenvalues of this operator and each one in
the diagonalized form has three different eigenvalues $E_1$, $E_2$, $E_3$.
Three different eigenfunctions $\psi _{\pm }^{\left( \pm \right) ,(3)}$
correspond to these eigenvalues of the spectrum. But the operator $\lambda ^3
$ does not commute with (1.7) and consequently, is not conserved operator
inspite of the field (1.4) directed along the third axis in the color space
as well. This is a reason, why $\lambda ^3$ does not determine the branches
of the energy spectum (1.9) and the projection of this operator onto the
chromomagnetic field is not suitable quantity for the description of color
states. Projection of $\overrightarrow{\lambda }$ operator onto the color
vector $\overrightarrow{I}$, i.e. the scalar product (1.27) commute with the
Hamiltonian (1.7), i.e. is a conserved quantity. As we see from this
analysis, the term with $\pm $ sign in the energy spectra (1.9) coming from
the$\left( \lambda ^bI^b\right) _{\pm }$ term in the Hamoltonians (1.6) and,
in fact, resutls to splitting of the energy spectrum into three branches ,
which correspond to the eigenvalues of this projection. The wave functions $%
\psi _{\pm }^{\left( \pm \right) }$ describing states with the energies from
these branches of spectra, are eigenfunctions of the operator (1.27) as
well. Though the two different operators $\left( \lambda ^bI^b\right) _{-}$
and $\left( \lambda ^bI^b\right) _{+}$ correspond to the two different
eigenvalues of the spin operator, the different eigenvalues of these
operators coincide: $\left( \lambda ^bI^b\right) _{+}^{(+)}=\left( \lambda
^bI^b\right) _{-}^{(-)}=-{\cal GP}_{\bot }$, $\left( \lambda ^bI^b\right)
_{+}^{(-)}=\left( \lambda ^bI^b\right) _{-}^{(+)}={\cal GP}_{\bot }$, $%
\left( \lambda ^bI^b\right) _{+}^{(3)}=\left( \lambda ^bI^b\right)
_{-}^{(3)}=0$. Consequently, the wave functions of the states with the same
spin projection, but with the different value of $\left( \lambda
^bI^b\right) ^r$ describe the different energy branches: $\psi _{+}^{\left(
+\right) }\rightarrow E_2$, $\psi _{+}^{\left( -\right) }\rightarrow E_1$
and $\psi _{-}^{\left( +\right) }\rightarrow E_1$, $\psi _{-}^{\left(
-\right) }\rightarrow E_2$ and wave functions of the states with the same
projection $\left( \lambda ^bI^b\right) ^r$, but with the different spin
projection describe the different energy branches as well: $\psi
_{+}^{\left( +\right) }\rightarrow E_2$, $\psi _{-}^{\left( +\right)
}\rightarrow E_1$ and $\psi _{+}^{\left( -\right) }\rightarrow E_1$, $\psi
_{-}^{\left( -\right) }\rightarrow E_2$. Since this operator has three
eigenvalues instead of six, it arise twofold degeneracy, which cannot be
classifed either spin degeneracy or degeneracy on eigenvalues of projection $%
\left( \lambda ^bI^b\right) ^r$. In the transformed space, where the
operator $\left( \lambda ^bI^b\right) $ has got its diagonal form: 
\begin{equation}
\label{1.28}\left( \lambda ^bI^b\right) ^r=\left( 
\begin{array}{cccccc}
-{\cal GP}_{\bot } & 0 & 0 & 0 & 0 & 0 \\ 
0 & {\cal GP}_{\bot } & 0 & 0 & 0 & 0 \\ 
0 & 0 & 0 & 0 & 0 & 0 \\ 
0 & 0 & 0 & {\cal GP}_{\bot } & 0 & 0 \\ 
0 & 0 & 0 & 0 & -{\cal GP}_{\bot } & 0 \\ 
0 & 0 & 0 & 0 & 0 & 0
\end{array}
\right) ,
\end{equation}
it can be written as the product of two operators $-\sigma ^3\left( \lambda
^bI^b\right) _{+}^r$ and easily found that degeneracy of the energy spectrum
is degeneracy on this product. Having expressed the spectrum branches in
terms of the projections of $\sigma ^3$ and $\left( \lambda ^bI^b\right)
_{+}^r$ we can establish the values of these projections, on which these
branches coincide:%
$$
\begin{array}{c}
E\left( \sigma ^3=1;\ \left( \lambda ^bI^b\right) ^r=-
{\cal GP}_{\bot }\right) =E\left( \sigma ^3=-1;\ \left( \lambda ^bI^b\right)
^r={\cal GP}_{\bot }\right) =E_2, \\ E\left( \sigma ^3=1;\ \left( \lambda
^bI^b\right) ^r={\cal GP}_{\bot }\right) =E\left( \sigma ^3=-1;\ \left(
\lambda ^bI^b\right) ^r=-{\cal GP}_{\bot }\right) =E_1.
\end{array}
$$
Thus we conclude, that this is degeneracy on the quantity (1.27), which
include spin, color spin and momentum, inspite there is no degeneracy on
this quantities separately. In the result of this degeneracy the states
having different quantum numbers $\sigma ^3$ and $\left( \lambda
^bI^b\right) ^r$ have got the same energy: $\psi _{+}^{\left( +\right) }$, $%
\psi _{-}^{\left( -\right) }\rightarrow E_2$ and $\psi _{-}^{\left( +\right)
}$, $\psi _{+}^{\left( -\right) }\rightarrow E_1$. After this analysis we
come to idea to write the energy states as eigenfunctions of $\left( \lambda
^bI^b\right) _{\pm }^r$ operator, introducing unit eigenvectors $\zeta $$%
^{\left( \pm \right) }$ of these operators: 
\begin{equation}
\label{1.29}\left( \lambda ^bI^b\right) _{\pm }^r\zeta ^{\left( \pm \right)
}=\pm {\cal GP}_{\bot }\lambda ^3\zeta ^{\left( \pm \right) }=\pm {\cal GP}%
_{\bot }\left( \pm \zeta ^{\left( \pm \right) }\right) ,\quad \left( \lambda
^bI^b\right) _{+}^r\zeta ^{\left( 3\right) }={\cal GP}_{\bot }\lambda
^3\zeta ^{\left( 3\right) }=0,
\end{equation}
which are basis vectors of the transformed color space as well:%
$$
\ \zeta ^{(+)}=\frac 1{\sqrt{3}}\left( 
\begin{array}{c}
1 \\ 
0 \\ 
0
\end{array}
\right) ,\ \zeta ^{(-)}=\frac 1{\sqrt{3}}\left( 
\begin{array}{c}
0 \\ 
1 \\ 
0
\end{array}
\right) ,\ \zeta ^{(3)}=\frac 1{\sqrt{3}}\left( 
\begin{array}{c}
0 \\ 
0 \\ 
1
\end{array}
\right) . 
$$
Then, the energy states can be expressed on this basis as following: 
\begin{equation}
\label{1.30}\psi _{\pm }^{\left( \pm \right) }=\zeta ^{(\pm )}\xi _{\pm
}F(r).
\end{equation}

As we noted, the term $\left( \lambda ^bI^b\right) $ describes the
interaction of the particle with the external field. This interaction occur
by means of chromomagnetic moment of the particle due to its spin, which is
included in the last term in (1.27) and by the chromomagnetic moment
acquired due to orbital moment, which we suppose, hidden in the first term
of (1.27). In the abelian theory, in motion of electron in the axial
magnetic field given by vector potential $A_\mu =\left( 0,\ -\frac 12yH_z,\
\frac 12xH_z,\ 0,\right) $ the term $A_jp_j$ in the Hamiltonian is written
as the interaction term of the magnetic moment of electron due to orbital
moment $L_z$ with the magnetic field $H_z$ [7] 
$$
\frac e{m_0c}A_jp_j=-\frac{eH_z}{2m_0c}L_z=-\frac{eH_z\hbar }{2m_0c}m. 
$$
This term eliminates the degeneracy of energy levels on magnetic quantum
number $m$ and leads to physical effect of splitting of energy spectrum
levels in the magnetic field. In our non-abelian Landau problem the term $%
g\lambda ^bA_j^bp_j=2gA_jp_j$ cannot be written proportional to the orbital
moment. It has color matrix structure%
$$
\lambda ^bA_j^bp_j={\cal G}\left( 
\begin{array}{ccc}
0 & p_{-} & 0 \\ 
p_{+} & 0 & 0 \\ 
0 & 0 & 0 
\end{array}
\right) 
$$
and acting of the $p_{\pm }$ on the wave functions $e^{im\varphi }J_m\left(
p_{\bot }r\right) $ is given by (1.22). In quantized spectrum case $p_{\bot
}=\alpha _m^{(N)}/r_0$ and for this interaction term we have following
eigenvalues:%
$$
\lambda ^bA_j^bp_je^{im\varphi }J_m\left( p_{\bot }r\right) \left( 
\begin{array}{c}
\xi _{\pm }^{(1)} \\ 
\xi _{\pm }^{(2)} \\ 
\xi _{\pm }^{(3)} 
\end{array}
\right) ={\cal G}\frac{\alpha _m^{(N)}}{r_0}\left( 
\begin{array}{c}
-e^{i\left( m-1\right) m\varphi }J_{m-1}\left( \alpha _m^{(N)}r/r_0\right)
\xi _{\pm }^{(2)} \\ 
\;e^{i\left( m+1\right) m\varphi }J_{m+1}\left( \alpha _m^{(N)}r/r_0\right)
\xi _{\pm }^{(1)} \\ 
0 
\end{array}
\right) . 
$$
As is seen, the term $g\lambda ^bA_j^bp_j$ in Hamiltonian does not lead to
appearance of the terms with factor of chromomagnetic quantum number $m$ and
shifts the state with $m$ to the states with $m\pm 1$. But this term
contains $\alpha _m^{(N)},$ which depends on $m$ and on $N$. This means,
that the term $g\lambda ^bA_j^bp_j$ splits the spectrum into series defined
by $N$ and levels in series are defined by $m$. Let us remind, that the
radia $a_m^{(N)}$ of orbits, in which rotates the particle in this field, is
also determined by $\alpha _m^{(N)}$, i.e. by both quantum numbers $m$ and $%
N $ [4]: $a_m^{(N)}=mr_0/\alpha _m^{(N)}$. Thus we conclude, that the
interaction term $gA_jp_j$ in non-abelian theory, as a result of the
interaction of chromomagnetic moment due to the orbital moment with the
chromomagnetic background, eliminates degeneracy on $m$ and $N$ and split
the energy levels in the quantized spectrum case. Beside splitting, this
interaction shifts particle from the state $m$ to the states $m\pm 1$ and
consequently, from the orbit $a_m^{(N)}$ to the orbits $a_{m\pm 1}^{(N)}$ in
dependence of projection of $\lambda ^3$ onto the chromomagnetic field. In
the transformed color space, according to (1.39) this interaction term
splits spectrum into branches, and each branch contains factor ${\cal G}%
P_{\bot }={\cal G}\sqrt{p_{\bot }^2+{\cal G}^2/4}={\cal G}\sqrt{\left(
\alpha _m^{(N)}/r_0\right) ^2+{\cal G}^2/4}$, which splits spectrum banches
into series.

\section{Spherical background}

\setcounter{equation}{0}

Let us consider the case of constant spherical chromomagnetic field defined
by the constant vector potential: 
\begin{equation}
\label{2.1}A_1^a=\sqrt{\tau }\delta _{1a},\ A_2^a=\sqrt{\tau }\delta _{2a},\
A_3^a=\sqrt{\tau }\delta _{3a},\ A_0^a=0. 
\end{equation}
This field has the strength tensor with the following non-zero components 
\begin{equation}
\label{2.2}F_{23}^1=F_{31}^2=F_{12}^3=H_x^1=-H_y^2=H_z^3=g\tau 
\end{equation}
and possesses spherical symmetry in ordinary space and in subspace of first
three coordinates in the color space. Squared Dirac equation (1.2) in the
field (2.1) has got the following explicit form: 
\begin{equation}
\label{2.3}\left( {\bf p}^2+M^2+\frac{3g^2\tau }4+g\tau ^{1/2}\lambda ^ap^a- 
\frac{g^2\tau }2\sigma ^a\lambda ^a\right) \Psi =E^2\Psi . 
\end{equation}
The spinor $\Psi $ has two components $\psi _{+}$ and $\psi _{-}$, and each
of them transforms under fundamental representation of $SU_c(3)$ color
group. Because the external field (2.2) has the three non-zero components in
the both spaces, in this case the spin and color spin states of the particle
cannot be described by the projection of the $\sigma ^3$ and $\lambda ^3$
operators. So, the components $\psi _{\pm }^{\left( a\right) }$ are not
eigenfunctions of these operators and do not describe the states with
definite value of their projection onto the field.

Equation (2.3) contains the non-diagonal $\sigma ^a$ and $\lambda ^a$
matrices, which make it non-diagonal. As a result, this equation does not
split into independent ones for the components $\psi _{\pm }^{\left(
a\right) }$. From (2.3) it can be obtained the determinant equation, that is
same for all states $\psi _j^{(i)}$ $\left( i=1,2\right) $ and has following
solution [4]: 
\begin{equation}
\label{2.4}\psi _j^{(i)}\left( {\bf r}\right) =\frac{C_l^\nu }{\mid {\bf p}%
\mid \sqrt{r}}J_{l+1/2}\left( \mid {\bf p}\mid r\right) Y_l^m\left( \theta
,\varphi \right) \xi _j^{(i)}\ , 
\end{equation}
where $J_{l+1/2}\left( \mid {\bf p}\mid r\right) $ is the first kind Bessel
function, $Y_l^m\left( \theta ,\varphi \right) $ are spheric functions, $%
C_l^\nu $ are normalizing constants $\left( C_l^\nu =\mid {\bf p}\mid
\right) $ and $\xi _j^{(i)}$ are the spin and color spin parts of the wave
function. Solving (2.3) for the energy gives the following branches of
continuous spectrum in the infinite motion case [10], [11]: 
$$
E_{1,2}^2=\left( \sqrt{{\bf p}^2}\mp \frac{{\cal G}}2\right) ^2+M^2,\
E_{3,4}^2=\left( \sqrt{{\bf p}^2+{\cal G}^2}\mp \frac{{\cal G}}2\right)
^2+M^2,\ E_5^2={\bf p}^2+M^2, 
$$
and the quantized spectrum branches in the case of the finite motion inside
sphere [4]: 
\begin{equation}
\label{2.5}E_{1,2}^2=\left( \frac{\alpha _l^{(N)}}{r_0}\mp \frac{{\cal G}}%
2\right) ^2+M^2,\ E_{3,4}^2=\left( \sqrt{\left( \frac{\alpha _l^{(N)}}{r_0}%
\right) ^2+{\cal G}^2}\mp \frac{{\cal G}}2\right) ^2+M^2,\ E_5^2=\left( 
\frac{\alpha _l^{(N)}}{r_0}\right) ^2+M^2. 
\end{equation}
Here $\alpha _l^{(N)}$ are zeros of $J_{l+1/2}\left( \mid {\bf p}\mid
r\right) $, $r_0$ is the radius of sphere and $l$ is the orbital quantum
number. Quantized and continuous spectra are interrelated by means of
replacement $\mid {\bf p}\mid =\alpha _l^{(N)}/r_0$, which is same to
quantization of momentum in standing waves. Motion in this case takes place
on $s$ $p$ $d$ $f$ ... orbitals. As we observe, the additional non-zero
components of the field strength tensor $F_{23}^1$ and $F_{31}^2$ apparently
break the remained color and spin symmetry in Hamiltonian. In the result of
this breakdown the spectrum splits into four branches as distinct from the
axial background field case, i.e. lead to elimination of the degeneracy of
the energy spectrum. From the spectrum (2.5) is seen also, that energy $E_5$
does not contain the interaction term with the external field and so, does
not split into branches. The state $\psi _j^{(3)}\left( {\bf r}\right) $
like to state of the spinless and colorless particle. But solving (2.3) by
means of its determinant leads to the absence of correspondence between the
branches of spectrum (2.5) and the solutions (2.4). Consequently, it is
reasonable to find the wave functions, which will describe the states having
definite energy from these branches, i.e. to find energy states in the field
(2.1).

Let us write the explicit matrix form of Hamiltonian for (2.3) in the
spin-color spin space:

\begin{equation}
\label{2.6}H=\left( 
\begin{array}{cccccc}
h_{11} & {\cal G}p_{-} & 0 & 0 & 0 & 0 \\ 
{\cal G}p_{+} & h_{22} & 0 & -{\cal G}^2 & 0 & 0 \\ 
0 & 0 & h_{33} & 0 & 0 & 0 \\ 
0 & -{\cal G}^2 & 0 & h_{44} & {\cal G}p_{-} & 0 \\ 
0 & 0 & 0 & {\cal G}p_{-} & h_{55} & 0 \\ 
0 & 0 & 0 & 0 & 0 & h_{66}
\end{array}
\right) .
\end{equation}
Diagonal elements $h_{ii}$ have the expressions:

\begin{equation}
\label{2.7}
\begin{array}{c}
h_{11}=
{\cal P}^2+{\cal G}^2/4+{\cal G}p_3\;\; \\ h_{22}=
{\cal P}^2+5{\cal G}^2/4-{\cal G}p_3 \\ h_{33}=
{\cal P}^2\;\;\;\;\;\;\;\;\;\;\;\;\;\;\;\;\;\;\;\;\;\;\; \\ h_{44}=
{\cal P}^2+5{\cal G}^2/4+{\cal G}p_3 \\ h_{55}=
{\cal P}^2+{\cal G}^2/4-{\cal G}p_3\;\; \\ h_{66}={\cal P}%
^2\;\;\;\;\;\;\;\;\;\;\;\;\;\;\;\;\;\;\;\;\;\;\;\;
\end{array}
,
\end{equation}
where ${\cal P}^2={\bf p}^2+M^2$. The Hamiltonian (2.6) has not
quasidiagonal form and so, we cannot diagonalize separately two matrices $%
H_{\pm }$ with dimension $3\times 3$ instead of one with dimension $6\times 6
$. In order to diagonalize (2.6) we should make transformation in the
combined space of spin and color spin. We can find this transformation and
diagonal form of (2.6) applying the method used in the previous section.

As was asserted in the previous section, the diagonal form of $H^{\prime }$
is unique and transformation $U$ transforming $H$ into $H^{\prime }$ is
unique and unitary due hermicity of $H$. Relying on these two properties, we
can find the matrices $H^{\prime }$ and ${\cal U}$ simultaneously. From the
beginning we find the determinant of this matrix, that equals to

\begin{equation}
\label{2.8}\det H=\left( {\cal P}^2\right) ^2\left( \left( {\cal P}^2+{\cal G%
}^2/4\right) ^2-{\cal G}^2p^2\right) \left( \left( {\cal P}^2+5{\cal G}%
^2/4\right) ^2-{\cal G}^2p^{\prime 2}\right) =\det H^{\prime }
\end{equation}
and can be written as a product of six factors $f_i$:%
$$
f_{1,2}={\cal P}^2,\ f_{3,4}={\cal P}^2+{\cal G}^2/4\pm {\cal G}p,\ f_{5,6}=%
{\cal P}^2+5{\cal G}^2/4\pm {\cal G}p^{\prime }, 
$$
where $p=\mid {\bf p}\mid =\sqrt{{\bf p}^2},\ p^{\prime }=\sqrt{{\bf p}^2+%
{\cal G}^2}.$ Sum of these factors equals to the trace of the Hamiltonian
(2.6):%
$$
\sum\limits_if_i=6{\cal P}^2+3{\cal G}^2=\tr H=\sum\limits_ih_{ii}=\sum%
\limits_ih_{ii}^{\prime }. 
$$
According to the invariance of the trace and determinant of the matrix under
the unitary transformation, we may suppose that, the factors $f_i$ are the
diagonal elements of diagonalized Hamiltonian $H^{\prime }$. But this is not
enough in order to find the explicit form of $H^{\prime }$, as we do not
know the place of each of factors $f_i$ along the diagonal of $H^{\prime }$
and have no rule to determine this place. At first, we can assume some
identification of factors $f_i$ with the diagonal elements of $H^{\prime }$
and then write the equation (1.13) explicitly for this assumption with
unknown ${\cal U}$. Equations for $u_{ij}$ obtained from (1.13) together
with the equations obtained from unitarity condition will be solved
regularly, if diagonal elements were identified properly. If the factors $f_i
$ were corresponded to the diagonal elements $h_{ii}^{\prime }$ incorrectly,
then solving (1.13) will lead to mathematical nonsense due to uniqueness of $%
H^{\prime }$. So we make the following identification: 
\begin{equation}
\label{2.9}
\begin{array}{c}
h_{11}^{\prime }\equiv f_4=
{\cal P}^2+g^2\tau /4-{\cal G}p\;\;\; \\ h_{22}^{\prime }\equiv f_6=
{\cal P}^2+5g^2\tau /4-{\cal G}p^{\prime } \\ h_{33}^{\prime }\equiv f_1=
{\cal P}^2\;\;\;\;\;\;\;\;\;\;\;\;\;\;\;\;\;\;\;\;\;\;\;\; \\ h_{44}^{\prime
}\equiv f_5=
{\cal P}^2+5g^2\tau /4+{\cal G}p^{\prime } \\ h_{55}^{\prime }\equiv f_3=
{\cal P}^2+g^2\tau /4+{\cal G}p\;\;\; \\ h_{66}^{\prime }\equiv f_2={\cal P}%
^2\;\;\;\;\;\;\;\;\;\;\;\;\;\;\;\;\;\;\;\;\;\;\;\;
\end{array}
.
\end{equation}
Writing (1.13) for this choice, we get systems of linear equations relating
elements $u_{ij}$ of transformation matrix ${\cal U}$, which are simplified
into following form:

$$
\left\{ 
\begin{array}{c}
u_{11}=u_{21}p_{-}/\left( p-p_3\right) \\ 
u_{41}=u_{21}\;\;\;\;\;\;\;\;\;\;\;\;\;\;\;\;\;\;\;\; \\ 
u_{51}=u_{21}p_{+}/\left( p+p_3\right) 
\end{array}
\right. \qquad \left\{ 
\begin{array}{c}
u_{12}=-u_{22}p_{-}/\left( p^{\prime }+p_3- 
{\cal G}\right) \;\;\;\;\;\;\;\;\;\;\;\;\;\;\;\;\;\; \\ u_{42}=-u_{22}\left(
p^{\prime }-p_3- 
{\cal G}\right) /\left( p^{\prime }+p_3-{\cal G}\right) \\ 
u_{52}=-u_{22}p_{+}/\left( p^{\prime }+p_3-{\cal G}\right)
\;\;\;\;\;\;\;\;\;\;\;\;\;\;\;\;\;\; 
\end{array}
\right. 
$$
\begin{equation}
\label{2.10}\left\{ 
\begin{array}{c}
u_{14}=u_{24}p_{-}/\left( p^{\prime }-p_3+ 
{\cal G}\right) \;\;\;\;\;\;\;\;\;\;\;\;\;\;\;\;\;\;\;\;\; \\ 
u_{44}=-u_{24}\left( p^{\prime }+p_3+ 
{\cal G}\right) /\left( p^{\prime }-p_3+{\cal G}\right) \\ 
u_{54}=-u_{24}p_{+}/\left( p^{\prime }-p_3+{\cal G}\right)
\;\;\;\;\;\;\;\;\;\;\;\;\;\;\;\;\;\; 
\end{array}
\right. \qquad \ \left\{ 
\begin{array}{c}
u_{15}=-u_{25}p_{-}/\left( p+p_3\right) \\ 
u_{45}=u_{25}\;\;\;\;\;\;\;\;\;\;\;\;\;\;\;\;\;\;\;\;\;\; \\ 
u_{55}=-u_{25}p_{+}/\left( p-p_3\right) 
\end{array}
\right. 
\end{equation}
All other $u_{ij}$ are zero except for $u_{33}$, $u_{36}$, $u_{63}$ and $%
u_{66}$ elements.

All other elements $u_{ij}$ in (2.10) are expressed in terms of $u_{2j}$.
Equations (2.10) relates 32 unknowns by 24 relations. So, in order to find
all $u_{ij}$ we need in additional relations. For this purpose we can use
relations of unitarity condition, which are 6 equations $\sum%
\limits_{j=1}^6u_{ij}u_{ij}^{*}=1$ and 30\ ones\ $\
\sum\limits_{j=1}^6u_{ij}u_{kj}^{*}=0$, but these relations are nonlinear.
Besides, some of the unitarity relations coincide with another ones after
taking into account relations of (2.10) in them. Thus, taking into account
the relations of (2.10) in the unitary relations and solving them we find
only the module of $u_{2i}$: 
$$
u_{21}u_{21}^{*}=\frac{p_{\bot }^2}{4p^2},\qquad u_{22}u_{22}^{*}=\frac{%
\left( p^{\prime }+p_3-{\cal G}\right) ^2}{4p^{\prime }\left( p^{\prime }-%
{\cal G}\right) }, 
$$

\begin{equation}
\label{2.11}u_{24}u_{24}^{*}=\frac{\left( p^{\prime }+p_3-{\cal G}\right) ^2
}{4p^{\prime }\left( p^{\prime }+{\cal G}\right) },\qquad u_{25}u_{25}^{*}=
\frac{p_{\bot }^2}{4p^2}.
\end{equation}
The arguments of $u_{2i}$ remain unknown and we parametrize them introducing
free angles $\alpha ,\beta ,\gamma $ and $\delta $ as below: 
$$
u_{21}=\frac{p_{\bot }}{2p}e^{i\alpha },\quad u_{22}=\frac{p^{\prime }+p_3-%
{\cal G}}{\left[ 4p^{\prime }\left( p^{\prime }-{\cal G}\right) \right]
^{1/2}}e^{i\beta }, 
$$
\begin{equation}
\label{2.12}u_{24}=\frac{p^{\prime }+p_3-{\cal G}}{\left[ 4p^{\prime }\left(
p^{\prime }+{\cal G}\right) \right] ^{1/2}}e^{i\gamma },\quad u_{25}=\frac{%
p_{\bot }}{2p}e^{i\delta }.
\end{equation}
In the system (2.10) there is no equation relating these four elements and
so, we cannot express one angle in terms of another ones%
\footnote{This was made in the previous case.}. As elements $u_{3j}$, $u_{j3}
$, $u_{6j}$ and $u_{j6}$ equal to zero, the equation (1.13) gives trival
relations for the elements $u_{33},u_{36},u_{63}$ and $u_{66}$ and does not
relate them with another $u_{ij}$ in (2.10). The relations for these
elements, obtained from the unitarity condition, consist in the following
ones: 
\begin{equation}
\label{2.13}\left\{ 
\begin{array}{c}
u_{33}u_{33}^{*}+u_{36}u_{36}^{*}=1 \\ 
u_{63}u_{63}^{*}+u_{66}u_{66}^{*}=1 \\ 
u_{33}u_{63}^{*}+u_{36}u_{66}^{*}=0
\end{array}
\right. .
\end{equation}
The relations in (2.13) are nonlinear and their number less than the number
of unknowns. Consequently, we get one more free parameter introducing angle $%
\theta $ in order to parametrize elements in (2.13) as below:%
$$
u_{33}=\cos \theta ,\quad u_{36}=\sin \theta ,\quad u_{63}=\sin \theta
,\quad u_{66}=-\cos \theta . 
$$
It is reasonable to assume, that under ${\cal U}$ transformation the states $%
\psi _{\pm }^{\left( 3\right) }$do not change, since these states like
colorless states $\left( \lambda ^3=0\right) $. Under this assumption we can
fix parameter $\theta =0$ and then $u_{33}=-u_{66}=1,$ $u_{36}=u_{63}=0.$

As has been expressed before, the Hamiltonian (2.6) owing to non-zero $h_{42}
$ and $h_{24}$ elements has not a quasidiagonal form and mixes different
spin states. So, we cannot demand the unimodularity of any block of the $%
{\cal U}$ matrix, as we made in the previous case. We also have no
additional conditions to cut down the number of free parameters ($\alpha
,\beta ,\gamma ,\delta )$ or fix them. Thus, the ${\cal U}$ matrix contains
four free parameters and has the following explicit form: 
\begin{equation}
\label{2.14}{\cal U}=\left( 
\begin{array}{cccccc}
p_{-}{\cal P}_{\bot }/{\cal P}_3 & -p_{-}/{\cal P}_1 & 0 & p_{-}/{\cal P}_2
& p_{-}{\cal P}_{\bot }^{\prime }/{\cal P}_4 & 0 \\ 
{\cal P}_{\bot } & \left( p^{\prime }+p_3-{\cal G}\right) /{\cal P}_1 & 0 & 
\left( p^{\prime }-p_3+{\cal G}\right) /{\cal P}_2 & {\cal P}_{\bot
}^{\prime } & 0 \\ 
0 & 0 & 1 & 0 & 0 & 0 \\ 
{\cal P}_{\bot } & \left( p^{\prime }-p_3-{\cal G}\right) /{\cal P}_1 & 0 & 
-\left( p^{\prime }+p_3+{\cal G}\right) /{\cal P}_2 & {\cal P}_{\bot
}^{\prime } & 0 \\ 
p_{+}{\cal P}_{\bot }/{\cal P}_4 & p_{+}/{\cal P}_1 & 0 & -p_{+}/{\cal P}_2
& p_{+}{\cal P}_{\bot }^{\prime }/{\cal P}_3 & 0 \\ 
0 & 0 & 0 & 0 & 0 & -1
\end{array}
\right) ,
\end{equation}
where ${\cal P}_1=\left[ 4p^{\prime }\left( p^{\prime }-{\cal G}\right)
\right] ^{1/2}e^{-i\beta }$, ${\cal P}_2=\left[ 4p^{\prime }\left( p^{\prime
}+{\cal G}\right) \right] ^{1/2}e^{-i\gamma }$, ${\cal P}_{3,4}=p\mp p_3$, $%
{\cal P}_{\bot }=\left( p_{\bot }/2p\right) e^{i\alpha }$, ${\cal P}_{\bot
}^{\prime }=-\left( p_{\bot }/2p\right) e^{i\delta }.$ Transforming the
basis vectors $\psi _j^{(i)}$ under the ${\cal U}$ transformation, we can
construct the energy states according to (1.20). Multiplying the ${\cal U}$
matrix (2.14) by the column of solutions (2.4) we find the energy states:%
$$
\psi _{+}^{(+)}=\frac{p_{-}{\cal P}_{\bot }}{{\cal P}_3}\psi _1^{(1)}\left( 
{\bf r}\right) -\frac{p_{-}}{{\cal P}_1}\psi _1^{(2)}\left( {\bf r}\right) +
\frac{p_{-}}{{\cal P}_2}\psi _2^{(1)}\left( {\bf r}\right) +\frac{p_{-}{\cal %
P}_{\bot }^{\prime }}{{\cal P}_4}\psi _2^{(2)}\left( {\bf r}\right) , 
$$
$$
\psi _{+}^{\left( -\right) }={\cal P}_{\bot }\psi _1^{(1)}\left( {\bf r}%
\right) +\frac{\left( p^{\prime }+p_3-{\cal G}\right) }{{\cal P}_1}\psi
_1^{(2)}\left( {\bf r}\right) +\frac{\left( p^{\prime }-p_3+{\cal G}\right) 
}{{\cal P}_2}\psi _2^{(1)}\left( {\bf r}\right) +{\cal P}_{\bot }^{\prime
}\psi _2^{(2)}\left( {\bf r}\right) , 
$$
$$
\psi _{-}^{\left( +\right) }={\cal P}_{\bot }\psi _1^{(1)}\left( {\bf r}%
\right) +\frac{\left( p^{\prime }-p_3-{\cal G}\right) }{{\cal P}_1}\psi
_1^{(2)}\left( {\bf r}\right) -\frac{\left( p^{\prime }+p_3+{\cal G}\right) 
}{{\cal P}_2}\psi _2^{(1)}\left( {\bf r}\right) +{\cal P}_{\bot }^{\prime
}\psi _2^{(2)}\left( {\bf r}\right) , 
$$
\begin{equation}
\label{2.15}\psi _{-}^{(-)}=\frac{p_{+}{\cal P}_{\bot }}{{\cal P}_4}\psi
_1^{(1)}\left( {\bf r}\right) +\frac{p_{+}}{{\cal P}_1}\psi _1^{(2)}\left( 
{\bf r}\right) -\frac{p_{+}}{{\cal P}_2}\psi _2^{(1)}\left( {\bf r}\right) +
\frac{p_{+}{\cal P}_{\bot }^{\prime }}{{\cal P}_3}\psi _2^{(2)}\left( {\bf r}%
\right) ,
\end{equation}
We can replace the all momentum operators in (2.15) by their eigenvalues.
This does not lead to any changes in (2.15). Remark, the transformation $%
{\cal U}$ acts on six dimensional combined space of spin and color spin, so
the up and down $\pm $ signs in $\psi _{\pm }^{(\pm )}$ in (2.15) are just
only notations and have no meanings of projection signs.

Comparing the spectrum branches (2.4) with the diagonal elements of $%
H^{\prime }$ (2.9), we write down obvious correspondence between them: 
$$
\begin{array}{c}
E_1\longrightarrow h_{11}^{\prime }, \\ 
E_2\longrightarrow h_{55}^{\prime }, \\ 
E_3\longrightarrow h_{22}^{\prime }, \\ 
E_4\longrightarrow h_{44}^{\prime }.
\end{array}
$$
Using this correspondence we can write the eigenvalue equation (1.10) for $%
H^{\prime }$ by eigenfunctions $\psi _{\pm }^{(\pm )}$ (2.15) and spectrum
(2.5):%
$$
\left( 
\begin{array}{cccccc}
h_{11}^{\prime } & 0 & 0 & 0 & 0 & 0 \\ 
0 & h_{22}^{\prime } & 0 & 0 & 0 & 0 \\ 
0 & 0 & h_{33}^{\prime } & 0 & 0 & 0 \\ 
0 & 0 & 0 & h_{44}^{\prime } & 0 & 0 \\ 
0 & 0 & 0 & 0 & h_{55}^{\prime } & 0 \\ 
0 & 0 & 0 & 0 & 0 & h_{66}^{\prime }
\end{array}
\right) \left( 
\begin{array}{c}
\psi _{+}^{\left( +\right) } \\ 
\psi _{+}^{\left( -\right) } \\ 
\psi _{+}^{(3)} \\ 
\psi _{-}^{\left( +\right) } \\ 
\psi _{-}^{\left( -\right) } \\ 
\psi _{-}^{(3)}
\end{array}
\right) =\left( 
\begin{array}{c}
E_1^2\psi _{+}^{\left( +\right) } \\ 
E_3^2\psi _{+}^{\left( -\right) } \\ 
E_5^2\psi _{+}^{(3)} \\ 
E_4^2\psi _{-}^{\left( +\right) } \\ 
E_2^2\psi _{-}^{\left( -\right) } \\ 
E_5^2\psi _{-}^{(3)}
\end{array}
\right) . 
$$
Thus, we find the wave functions of colored particle describing the states
having energy from the spectrum (2.4). This can be written in the following
correspondence between the states (2.15) and branches of the energy spectrum
(2.4):%
$$
\begin{array}{c}
E_1\longrightarrow \psi _{+}^{\left( +\right) }, \\ 
E_2\longrightarrow \psi _{-}^{\left( -\right) }, \\ 
E_3\longrightarrow \psi _{+}^{\left( -\right) }, \\ 
E_4\longrightarrow \psi _{-}^{\left( +\right) }.
\end{array}
$$
Consider now the operator, which determines the branches of spectra (2.4),
i.e. separate the term in Hamiltonian describing the interaction of the
particle with the external field. The non-diagonal part of Hamiltonian (2.6) 
\begin{equation}
\label{2.16}g\tau ^{1/2}\lambda ^b\left( p^b-\frac{g\tau ^{1/2}}2\sigma
^b\right) =g\lambda ^b\left( A_j^bp_j-\frac g4\epsilon
_{ijk}f^{acb}A_i^aA_j^c\sigma ^k\right) =\left( \lambda ^bI^b\right) 
\end{equation}
under the transformation (1.12) with the ${\cal U}$ matrix (2.14) transfoms
into the its diagonal form: 
\begin{equation}
\label{2.17}\left( \lambda ^bI^b\right) ^r={\cal U}^{-1}\left( \lambda
^bI^b\right) {\cal U}=\left( 
\begin{array}{cccccc}
-{\cal G}p & 0 & 0 & 0 & 0 & 0 \\ 
0 & -{\cal G}p^{\prime } & 0 & 0 & 0 & 0 \\ 
0 & 0 & 0 & 0 & 0 & 0 \\ 
0 & 0 & 0 & {\cal G}p^{\prime } & 0 & 0 \\ 
0 & 0 & 0 & 0 & {\cal G}p & 0 \\ 
0 & 0 & 0 & 0 & 0 & 0
\end{array}
\right) .
\end{equation}
Here $\epsilon _{ijk}$ is the unit antisymmetric tensor. As is seen from
(2.17), in this case the diagonal form of operator (2.17) has five different
eigenvalues, each one of which corresponds to the one branch of energy
spectrum (2.4). In non-transformed space this operator can also be written
in terms of the interactions of chromomagnetic moments with the
chromomagnetic background: 
\begin{equation}
\label{2.18}\left( \lambda ^bI^b\right) =g\lambda ^b\left( A_j^bp_j-\frac
12\epsilon _{ijk}F_{ij}^b\sigma ^k\right) =2g\left( A_jp_j-\frac 12\epsilon
_{ijk}F_{ij}\sigma ^k\right) .
\end{equation}
Second term in (2.18) describes the interaction of the particle with the
external field due to its spin and first one hides the interaction with the
external field due to orbital moment. In transformed space these two kind
interactions are joined in the eigenvalues $\pm $${\cal G}p$, $\pm {\cal G}%
p^{\prime }$ and $0$ and cannot be separated. The operator $\left( \lambda
^bI^b\right) $ commutes with the Hamiltonian (2.6) and the quantity
corresponding to it is conserved. Thus we conclude, that in external fields,
given by the noncommuting potentials (1.4) and (2.1), the projection $\left(
\lambda ^bI^b\right) $ describes the interaction of the chromomagnetic
moment of the particle with the external chromomagnetic field and cause the
splitting of the energy spectrum into the branches and levels. States with
definite energy in these fields are determined by this projection instead of
projection of spin and color spin onto the field, i.e. are defined in that
basis, in which this projection has got the diagonal form.

\section{Superpartner states}

\setcounter{equation}{0}

As we know [12,13], the equation (1.2) in the field (1.5) posseses
supersymmetry, i.e. for the motion with $p_3=0$ the Hamiltonian $H_1=H-M^2$
and two operators $Q_{\pm }=P_{\mp }a_{\pm }$ , which named the supercahrge
operators, form supersymmetry algebra in Quantum Mechanics [19-24]: 
\begin{equation}
\label{3.1}\left\{ Q_{+},Q_{-}\right\} =H_1,\ \left[ Q_{\pm },H_1\right]
=0,\ Q_{+}^2=Q_{-}^2=0. 
\end{equation}
Here the brace denotes anticommutator and the square bracket denotes
commutator. Defined by formulae

\begin{equation}
\label{3.2}P_{\pm }=P_1\pm iP_2,\ a_{\pm }=\frac 12\left( \sigma _1\pm
i\sigma _2\right) 
\end{equation}
the operators $P_{\pm }$ and$\ a_{\pm }$ obey the following commutation and
anticommutation relations for the annihilation and creation operators of
bosonic and fermionic states, respectively:

\begin{equation}
\label{3.3}\left[ P_{+},P_{-}\right] =\lambda ^3gH_z^3,\ \left\{
a_{+},a_{-}\right\} =1. 
\end{equation}
Let us remark, that the operators for creation and annihilation of the
bosonic states interchange roles for the state having value $\lambda ^3=1$
with the state having value $\lambda ^3=-1$ [13], that can be seen from
(3.3). Action of the fermionic operators $a_{\pm }$ turns over the spin of
particle: 
\begin{equation}
\label{3.4} 
\begin{array}{c}
a_{+}\xi _{-}=\left( 
\begin{array}{cc}
0 & 1 \\ 
0 & 0 
\end{array}
\right) \left( 
\begin{array}{c}
0 \\ 
1 
\end{array}
\right) =\left( 
\begin{array}{c}
1 \\ 
0 
\end{array}
\right) =\xi _{+},\quad a_{+}\xi _{+}=0; \\ 
a_{-}\xi _{+}=\left( 
\begin{array}{cc}
0 & 0 \\ 
1 & 0 
\end{array}
\right) \left( 
\begin{array}{c}
1 \\ 
0 
\end{array}
\right) =\left( 
\begin{array}{c}
0 \\ 
1 
\end{array}
\right) =\xi _{-},\quad a_{-}\xi _{-}=0. 
\end{array}
\end{equation}
Since the action of $a_{+}$ on $\left( 
\begin{array}{c}
\psi _{+} \\ 
\psi _{-} 
\end{array}
\right) $ gives $\psi _{-}$ and $a_{-}$ gives $\psi _{+}$ , the bosonic
creation operator $P_{+}$ acts only on the $\psi _{+}$ state and the
annihilation operator $P_{-}$ only on the $\psi _{-}$ state:%
$$
P_{+}\left( 
\begin{array}{c}
\psi _{+}^{(1)} \\ 
\psi _{+}^{(2)} \\ 
\psi _{+}^{(3)} 
\end{array}
\right) =\left( 
\begin{array}{ccc}
p_{+} & {\cal G} & 0 \\ 
0 & p_{+} & 0 \\ 
0 & 0 & p_{+} 
\end{array}
\right) \left( 
\begin{array}{c}
\psi _{+}^{(1)} \\ 
\psi _{+}^{(2)} \\ 
\psi _{+}^{(3)} 
\end{array}
\right) =\left( 
\begin{array}{c}
p_{+}\psi _{-}^{(1)}+ 
{\cal G}\psi _{-}^{(2)} \\ p_{+}\psi _{-}^{(2)} \\ 
p_{+}\psi _{-}^{(3)} 
\end{array}
\right) , 
$$
\begin{equation}
\label{3.5}P_{\_}\left( 
\begin{array}{c}
\psi _{-}^{(1)} \\ 
\psi _{-}^{(2)} \\ 
\psi _{-}^{(3)} 
\end{array}
\right) =\left( 
\begin{array}{ccc}
p_{-} & 0 & 0 \\ 
{\cal G} & p_{-} & 0 \\ 
0 & 0 & p_{-} 
\end{array}
\right) \left( 
\begin{array}{c}
\psi _{-}^{(1)} \\ 
\psi _{-}^{(2)} \\ 
\psi _{-}^{(3)} 
\end{array}
\right) =\left( 
\begin{array}{c}
p_{-}\psi _{+}^{(1)} \\ 
{\cal G}\psi _{+}^{(1)}+p_{-}\psi _{+}^{(2)} \\ p_{-}\psi _{+}^{(3)} 
\end{array}
\right) . 
\end{equation}
Using the supercharge operators it is easy to observe the spin diagonal form
of the Hamiltonian $H_1$:

\begin{equation}
\label{3.6}H_1\left( 
\begin{array}{c}
\psi _{+} \\ 
\psi _{-} 
\end{array}
\right) =\left( 
\begin{array}{c}
P_{-}P_{+}\qquad 0 \\ 
0\qquad P_{+}P_{-} 
\end{array}
\right) \left( 
\begin{array}{c}
\psi _{+} \\ 
\psi _{-} 
\end{array}
\right) . 
\end{equation}
Note, that action of the bosonic creation and annihilation operators $P_{\pm
}$ mix the different color states $\psi ^{(1)}$ and $\psi ^{(2)}$. In
Supersymmetric Quantum Mechanics each quantum mechanical state of particle
is labeled by fermionic and bosonic quantum numbers [19]-[21]. Superpartner
states in this theory are called the states having the same energy, but
different fermionic and bosonic quantum numbers. Action of the supercharge
operators $Q_{\pm }$ change the fermionic and bosonic quantum numbers and
convert the superpartner states into each other.

In the first section we have established that energy states having different
spin projection, namely the states $\psi _{+}^{\left( -\right) }$ and $\psi
_{-}^{\left( +\right) }$ and the states $\psi _{+}^{\left( +\right) }$ and $%
\psi _{-}^{\left( -\right) }$ have same energies. So, we have a reason to
look for superpartner states among the states $\psi _{+}^{\left( \pm \right)
}$ and $\psi _{-}^{\left( \pm \right) }$. Since these states differ by the
spin projection we can assert that they are turned into each other under the
action of the fermionic operators $a_{\pm }$. If we find a couple of
mutually hermitian bosonic operators $P_{\pm }$, which obey (3.3) and (3.6),
and change upper index of $\psi ^{\left( \pm \right) }$, i.e. acts by the
rule $P_{\pm }\psi ^{\left( \pm \right) }\rightarrow \psi ^{\left( \mp
\right) }$, then we can construct supercharge operators $Q_{\pm }$ by means
of these operators, that will obey the supersymmetry algebra (3.1) and will
map $\psi _{\pm }^{\left( \pm \right) }$ into each other:%
$$
Q_{+}\psi _{-}^{\left( +\right) }=q_1\psi _{+}^{\left( -\right) },\quad
Q_{+}\psi _{-}^{\left( -\right) }=q_2\psi _{+}^{\left( +\right) }; 
$$
\begin{equation}
\label{3.7}Q_{-}\psi _{+}^{\left( -\right) }=q_1^{\prime }\psi _{-}^{\left(
+\right) },\quad Q_{-}\psi _{+}^{\left( +\right) }=q_2^{\prime }\psi
_{-}^{\left( -\right) }. 
\end{equation}
The states $\psi _{+}^{\left( -\right) }$ and $\psi _{-}^{\left( +\right) }$
and the states $\psi _{+}^{\left( +\right) }$ and $\psi _{-}^{\left(
-\right) }$ will be superpartner states in the framework of this
supersymmetry. So, for the superpartnership of these states it is enough to
find suitable bosonic operators.

Under the ${\cal U}$ transformation the operators $P_{\pm }$ transform as
well: 
\begin{equation}
\label{3.8}U_{\pm }^{-1}H_{\pm }U_{\pm }=U_{\pm }^{-1}P_{\mp }P_{\pm }U_{\pm
}=U_{\pm }^{-1}P_{\mp }U_{\pm }U_{\pm }^{-1}P_{\pm }U_{\pm }=P_{\mp
}^{\prime }P_{\pm }^{\prime }. 
\end{equation}
Applying $U_{\pm }$ transformation we find $P_{\pm }^{\prime }$ operators in
the new basis: 
$$
P_{+}^{\prime }=\frac 1{2{\cal P}_{\bot }}\left( 
\begin{array}{ccc}
2p_{+}\left( {\cal P}_{\bot }+{\cal G}/2\right) & {\cal G}\left( {\cal P}%
_{\bot }+{\cal G}/2\right) e^{-2i\alpha } & 0 \\ 
-{\cal G}p_{+}^2\left( {\cal P}_{\bot }+{\cal G}/2\right) ^{-1}e^{2i\alpha }
& 2p_{+}\left( {\cal P}_{\bot }-{\cal G}/2\right) & 0 \\ 
0 & 0 & 2{\cal P}_{\bot }p_{+} 
\end{array}
\right) ,\qquad 
$$
\begin{equation}
\label{3.9}P_{-}^{\prime }=\frac 1{2{\cal P}_{\bot }}\left( 
\begin{array}{ccc}
2p_{-}\left( {\cal P}_{\bot }+{\cal G}/2\right) & -{\cal G}p_{-}^2\left( 
{\cal P}_{\bot }+{\cal G}/2\right) ^{-1}e^{-2i\alpha } & 0 \\ 
{\cal G}\left( {\cal P}_{\bot }+{\cal G}/2\right) e^{2i\alpha } & 
2p_{-}\left( {\cal P}_{\bot }-{\cal G}/2\right) & 0 \\ 
0 & 0 & 2{\cal P}_{\bot }p_{-} 
\end{array}
\right) .\qquad 
\end{equation}
Product of transformed bosonic operators are diagonalized Hamiltonians $%
H_{\pm }^{\prime }$ (1.17) and (1.26) with $M^2=0$ and $p_3=0$: 
\begin{equation}
\label{3.10}P_{-}^{\prime }P_{+}^{\prime }=H_{+}^{\prime },\quad
P_{+}^{\prime }P_{-}^{\prime }=H_{-}^{\prime }. 
\end{equation}
It is clear that, the new supercharge operators $Q_{\pm }^{\prime }=P_{\mp
}^{\prime }a_{\pm }$ will obey the supersymmetry algebra (3.1) with the
diagonalized Hamiltonian $H^{\prime }$: 
\begin{equation}
\label{3.11}\left\{ Q_{+}^{\prime },Q_{-}^{\prime }\right\} =H_1^{\prime },\
\left[ Q_{\pm }^{\prime },H_1^{\prime }\right] =0,\ Q_{+}^{\prime
2}=Q_{-}^{\prime 2}=0. 
\end{equation}
But, we see from the explicit form of $P_{\pm }^{\prime }$ the action of
these operators on $\Psi ^{\prime }=\left( 
\begin{array}{c}
\psi _{\pm }^{(+)} \\ 
\psi _{\pm }^{(-)} \\ 
\psi _{\pm }^{(3)} 
\end{array}
\right) $ does not maps the states $\psi ^{(+)}$ and $\psi ^{(-)}$ into each
other and mix these states. This means that, in the result of action of
these operators we get superposition of the $\psi ^{(\pm )}$ states, i.e
states having uncertain energy. So, the operators $P_{\pm }^{\prime }$ are
unusable to build supersymmetry algebra, in which the states $\psi ^{(+)}$
and $\psi ^{(-)}$ would be turned out superpartners. But, fortunately, it
can be constructed the another mutually hermitian conjugate couple of $%
P_{\pm }^{\prime \prime }$ operators%
$$
P_{+}^{\prime \prime }={\cal P}_{\bot }\lambda ^1+i\lambda ^2{\cal G}%
/2+p_{+}\left( I_3-I_2\right) =\left( 
\begin{array}{ccc}
0 & {\cal P}_{\bot }+{\cal G}/2 & 0 \\ 
{\cal P}_{\bot }-{\cal G}/2 & 0 & 0 \\ 
0 & 0 & p_{+} 
\end{array}
\right) , 
$$
\begin{equation}
\label{3.12}P_{-}^{\prime \prime }={\cal P}_{\bot }\lambda ^1-i\lambda ^2%
{\cal G}/2+p_{-}\left( I_3-I_2\right) =\left( 
\begin{array}{ccc}
0 & {\cal P}_{\bot }-{\cal G}/2 & 0 \\ 
{\cal P}_{\bot }+{\cal G}/2 & 0 & 0 \\ 
0 & 0 & p_{-} 
\end{array}
\right) , 
\end{equation}
which obey commutation relation $\left[ P_{+}^{\prime \prime },P_{-}^{\prime
\prime }\right] =2\lambda ^3{\cal GP}_{\bot }$. Product of these operators
also gives $H_{\pm }^{\prime }$: 
$$
P_{-}^{\prime \prime }P_{+}^{\prime \prime }=H_{+}^{\prime },\quad
P_{+}^{\prime \prime }P_{-}^{\prime \prime }=H_{-}^{\prime } 
$$
and so, the supercharge operators constructed using $Q_{\pm }^{\prime
}=P_{\mp }^{\prime \prime }a_{\pm }$ also obey the supersymmetry algebra
(3.11). Action of the $P_{\pm }^{\prime \prime }$ operators on $\Psi
^{\prime }$ change the upper index of the components of this wave function: 
$$
P_{+}^{\prime \prime }\left( 
\begin{array}{c}
\psi ^{(+)} \\ 
\psi ^{(-)} \\ 
\psi ^{(3)} 
\end{array}
\right) =\left( 
\begin{array}{c}
\left( 
{\cal P}_{\bot }+{\cal G}/2\right) \psi ^{(-)} \\ \left( 
{\cal P}_{\bot }-{\cal G}/2\right) \psi ^{(+)} \\ p_{+}\psi ^{(3)} 
\end{array}
\right) , 
$$
\begin{equation}
\label{3.13}P_{-}^{\prime \prime }\left( 
\begin{array}{c}
\psi ^{(-)} \\ 
\psi ^{(+)} \\ 
\psi ^{(3)} 
\end{array}
\right) =\left( 
\begin{array}{c}
\left( 
{\cal P}_{\bot }-{\cal G}/2\right) \psi ^{(+)} \\ \left( 
{\cal P}_{\bot }+{\cal G}/2\right) \psi ^{(-)} \\ p_{-}\psi ^{(3)} 
\end{array}
\right) . 
\end{equation}
Action of these operators on $\zeta ^{\left( \pm \right) ,3}$ is given by
the formulae:%
$$
P_{+}^{\prime \prime }\zeta ^{\left( +\right) }=\left( {\cal P}_{\bot }-%
{\cal G}/2\right) \zeta ^{\left( -\right) },\quad P_{+}^{\prime \prime
}\zeta ^{\left( -\right) }=\left( {\cal P}_{\bot }+{\cal G}/2\right) \zeta
^{\left( +\right) }; 
$$
$$
P_{-}^{\prime \prime }\zeta ^{\left( +\right) }=\left( {\cal P}_{\bot }+%
{\cal G}/2\right) \zeta ^{\left( -\right) },\quad P_{-}^{\prime \prime
}\zeta ^{\left( -\right) }=\left( {\cal P}_{\bot }-{\cal G}/2\right) \zeta
^{\left( +\right) }; 
$$
\begin{equation}
\label{3.14}P_{\pm }^{\prime \prime }\zeta ^{\left( 3\right) }=p_{\pm }\zeta
^{\left( 3\right) }. 
\end{equation}
So, this action changes the orientation of the $I^b$ vector in the
transformed color space. The operators $b_{\pm }=P_{\pm }^{\prime \prime
}/\left( 2{\cal GP}_{\bot }\right) ^{1/2}$ obey the Heisenberg-Weil algebra 
$$
\left[ b_{+},b_{-}\right] =1 
$$
for the state $\zeta ^{\left( +\right) }$, i.e. are the operators of
creation and annihilation, respectively. For the state $\zeta ^{\left(
-\right) }$ these operators obey the commutation relation%
$$
\left[ b_{-},b_{+}\right] =1, 
$$
that again means interchanging the roles of creation and annihilation
operators of $\zeta ^{\left( +\right) }$ state.

We can write the action of $Q_{\pm }^{\prime }$ operators on $\psi _{\pm
}^{(\pm )}$ using (3.14), (3.4) and (1.40) formulae: 
$$
Q_{+}^{\prime }\psi _{-}^{(+)}=P_{-}^{\prime \prime }a_{+}\xi _{-}\zeta
^{\left( +\right) }F\left( r\right) =\left( {\cal P}_{\bot }+{\cal G}%
/2\right) \psi _{+}^{\left( -\right) }, 
$$
$$
Q_{+}^{\prime }\psi _{-}^{(-)}=P_{-}^{\prime \prime }a_{+}\xi _{-}\zeta
^{\left( -\right) }F\left( r\right) =\left( {\cal P}_{\bot }-{\cal G}%
/2\right) \psi _{+}^{\left( +\right) }, 
$$
$$
Q_{-}^{\prime }\psi _{+}^{(-)}=P_{+}^{\prime \prime }a_{-}\xi _{+}\zeta
^{\left( -\right) }F\left( r\right) =\left( {\cal P}_{\bot }+{\cal G}%
/2\right) \psi _{-}^{\left( +\right) }, 
$$
\begin{equation}
\label{3.15}Q_{-}^{\prime }\psi _{+}^{(+)}=P_{+}^{\prime \prime }a_{-}\xi
_{+}\zeta ^{\left( +\right) }F\left( r\right) =\left( {\cal P}_{\bot }-{\cal %
G}/2\right) \psi _{-}^{\left( -\right) }, 
\end{equation}
$$
Q_{+}^{\prime }\psi _{+}^{(-)}=Q_{+}^{\prime }\psi _{+}^{(+)}=Q_{-}^{\prime
}\psi _{-}^{(+)}=Q_{-}^{\prime }\psi _{-}^{(-)}=0; 
$$
$$
Q_{-}^{\prime }\psi _{+}^{(3)}=p_{+}\psi _{-}^{(3)},\ Q_{+}^{\prime }\psi
_{-}^{(3)}=p_{-}\psi _{+}^{(3)}. 
$$
This action simultaneously changes the signs of the spin and $\left(
I^b\lambda ^b\right) $ projection so that the energy of the state remains
unchanged. The eigenvalues $q_i$ of the supercharge operators follow from
these actions:%
$$
q_1=q_1^{\prime }={\cal P}_{\bot }+{\cal G}/2,\quad q_2=q_2^{\prime }={\cal P%
}_{\bot }-{\cal G}/2. 
$$
If remind the relationship between the degree $n$ of the degeneracy of the
spectrum and the number $N$ of the supercharge operators $Q_{\pm }^{\prime }$%
:%
$$
n=2^{\left[ N/2\right] }, 
$$
where $\left[ N/2\right] $ denotes integer part of $N/2$, in this choice of
the supercharge operators it is easy to explain the twofold degeneracy of
the energy spectrum, which we have discussed in first section, as a result
of supersymmetry in the Hamiltonian%
\footnote{In [13,12,27] this degeneracy was related with the supersymmetry, but appropriate supercharge operators responsible for this degeneracy was not found, since the energy states was not determined.}%
. Thus, action of supercharge operators (3.15) ensure the superpartnership
of the energy states $\psi _{+}^{(-)}$ with $\psi _{-}^{(+)}$ and $\psi
_{+}^{(+)}$ with $\psi _{-}^{(-)}$.

Using (3.11) we can divide the Hamiltonian $H_1^{\prime }$ into bosonic $H_B$
and fermionic $H_F$ parts: 
\begin{equation}
\label{3.16} 
\begin{array}{c}
H_1^{\prime }=\left\{ Q_{+}^{\prime },Q_{-}^{\prime }\right\} =2 
{\cal GP}_{\bot }b_{+}b_{-}-2\lambda ^3{\cal GP}_{\bot }a_{+}a_{-}=2{\cal GP}%
_{\bot }\left( b_{+}b_{-}+\frac 12\right) + \\ 2{\cal GP}_{\bot }\left(
\left( -\lambda ^3\right) a_{+}a_{-}-\frac 12\right) =H_B+H_F. 
\end{array}
\end{equation}
The appearance of $\left( -\lambda ^3\right) $ factor in the fermionic part
is connected with different commutation rules for the bosonic operators of $%
\zeta ^{\left( -\right) }$ and $\zeta ^{\left( +\right) }$ states. Actually, 
$H_B$ also contains the term proportional to $\lambda ^3$. These two
Hamiltonians commute%
\footnote{$H_{B}$ and $H_{F}$  does not commute in another basis [27].} $%
\left[ H_B,H_F\right] =0$ and can be considered as Hamiltonians of two
independent oscillators having the same frequency $\omega =\left( 2{\cal GP}%
_{\bot }\right) ^{1/2}:$%
$$
H_B=\omega ^2\left( b_{+}b_{-}+\frac 12\right) ,\quad E_B=\omega ^2\left(
n_B+\frac 12\right) ; 
$$
\begin{equation}
\label{3.17}H_F=\omega ^2\left( \left( -\lambda ^3\right) a_{+}a_{-}-\frac
12\right) ,\quad E_B=\omega ^2\left( \left( -\lambda ^3\right) n_F-\frac
12\right) . 
\end{equation}
In respect to the supersymmetry each quantum mechanical state of the
particle with definite energy is described by the bosonic and fermionic
quantum numbers $n_B$ and $n_F$ , which accept values [19]: $n_B=0,1,2,3...;$
$n_F=0,1$. As is seen from (3.17) and (3.16), this takes place for $%
H_1^{\prime }$ and its eigenstates $\psi _{\pm }^{(\pm )}$ as well. Action
of the supercharge operators change these quantum numbers as following: 
\begin{equation}
\label{3.18} 
\begin{array}{c}
Q_{+}\left( n_B,n_F\right) =\left( n_B-1,n_F+1\right) ,\quad Q_{+}\left(
n_B-1,n_F+1\right) =0; \\ 
Q_{-}\left( n_B-1,n_F+1\right) =\left( n_B,n_F\right) ,\quad Q_{-}\left(
n_B,n_F\right) =0, 
\end{array}
\end{equation}
but total energy of bosonic and fermionic oscillators remain unchanged on
this action. If we label the states $\psi _{\pm }^{(\pm )}$ by these quantum
numbers as below:%
$$
\psi _{-}^{(-)},\psi _{-}^{(+)}\rightarrow \left( n_B,n_F\right) ;\quad \psi
_{+}^{(-)},\psi _{+}^{(+)}\rightarrow \left( n_B-1,n_F+1\right) 
$$
the two acting rules of the supercharge operators (3.18) and (3.15) will
agree.

\section{Discussion}

The non-diagonal generators of color group make the equations in Yang-Mills
Quantum Mechanics non-diagonal as well and we decided to diagonalize the
equation of motion for the case. Relying on hermicity of the Hamiltonian, we
find the unitary transformation, which diagonalizes this equation. Of
course, this transformation transforms the color structure of Hilbert space
and the transformed basis vectors are the eigenfunctions of the diagonal
Hamiltonian. This enabled us to establish the correspondence between the
eigenvalues of the Hamiltonian and its eigenvectors.

The diagonal form of the Hamiltonian turned up useful for the study of
supersummetry in the considered case. It allowed to construct the
supercharge operators mapping eigenvectors corresponding to the same
eigenvalues each into others. Superpartner property of the eigenvectors was
easily revealed on this formulation of supersymmetry. Also it became
possible to divide the diagonal Hamiltonian into two commuting parts
corresponding to the two oscillators of bosonic and fermionic states of
supersymmetry.

The number of papers [14] have been devoted to the study of supersymmetric
Yang-Mills Quantum Mechanics in connection with the conjecture of Banks,
Fischler, Shenker and Susskind concerning the equivalence of M-theory and
the $D=10$ supersymmetric Yang-Mills quantum mechanics [26]. We hope the
study of example of supersymmetry, which we have made here, will be useful
for the further study of supersymmetry in Yang -Mills Quantum Mechanics in
connection with the BFSS conjecture.

\vskip 1.00cm

{\large {\bf Acknowledgement}}

\vskip 0.4cm

The author acknowledges Prof. T.M. Aliev from METU (Ankara, Turkey) for
inviting him to participate in ''International Workshop on Physics Beyond
the Standard Model'' in 22-26 September, 2005 (Mugla, Turkey), where the
main part of the present paper was included in the talk of author.


\begin{thebibliography}{99}
\bibitem{1}  G.K.Savvidy, Phys. Lett. 71B (1979) 133, H.B. Nielsen and P.
Olesen, Nucl. Phys. B152, 75 (1979); N.K. Nielsen and P. Olesen, Nucl. Phys.
B144, 376 (1978)

\bibitem{2}  O. Nachtmann ''High Energy Collisions and Nonperturbative QCD''
in {\it Lectures\ on\ QCD. Applications}, ed. by F.L.H. Griesshammer and D.
Stoll, Springer (1997)

\bibitem{3}  V. I. Demchik and Skalozub, Eur. Phys. J. C 25, 291 (2002)
(arXiv: hep-ph/0110280); Eur. Phys. J. C 27, 601 (2003), (arXiv:
hep-ph/0208274)

\bibitem{4}  Sh. Mamedov, Eur. Phys. J.\ C 30, 583 (2003) (arXiv:
hep-ph/0306179)

\bibitem{5}  Sh. Mamedov, Eur. Phys. J.\ C 33, 537 (2004) (arXiv:
hep-ph/0310273)

\bibitem{6}  L. S. Brown and W. I. Weisberger, Nucl. Phys. B157, 285 (1979);
M. Reuter and C. Wetterich, Phys. Lett. B334, 412 (1994)

\bibitem{7}  A.A. Sokolov, I.M. Ternov and V. Ch. Zhukovskii, {\it \
Kvantovaya\ Mechanica}, M. Nauka (1979)

\bibitem{8}  H.J.W. M{\"u}ller-Kirsten {\it Introduction\ to\ Quantum\
Mechanics:\ Schr\"odinger\ Equation\ and\ Path\ Integral}, World Scientific
(2006)

\bibitem{9}  A. Bohm {\it Quantum\ Mechanics:\ Foundations\ and\ Applications%
}, Springer-Verlag (1986)

\bibitem{10}  Sh.S. Agaev, A.S. Vshivtsev, V.Ch. Zhukovsky, Yad. Fiz. 36
(1982) 1023; Sh.S. Agaev, A.S. Vshivtsev, V.Ch. Zhukovsky, O.F. Semyonov,
Sov.Phys.J.28 (1985) 29; Sh.S. Agaev, A.S. Vshivtsev, V.Ch. Zhukovsky, O.F.
Semyonov, Sov.Phys.J. 28 (1985) 67

\bibitem{11}  A. S. Vshivtsev, V. Ch. Zhukovskii, P. G. Midodashvili and A.
V. Tatarintsev, Izv. Vissh. Ucheb. Zav., Fizika 5 (Sov.Phys.J. N5) (1986) 47

\bibitem{12}  Sh. Mamedov, Jian-Zu Zhang and V. Ch. Zhukovskii, Mod. Phys.
Lett. A, vol.16, No 13 (2001) 845 (arXiv: hep/th-0101183)

\bibitem{13}  V. Ch. Zhukovskii, Sov. Phys. JETP 63 (4) (1986) 663

\bibitem{14}  J. Wosiek Nucl.Phys. B644 (2002) 85 (arXiv: hep-th/0203116);
M. Campostrini and J. Wosiek Nucl. Phys. B703 (2004) 454 (arXiv:
hep-th/0407021); Phys. Lett. B619 (2005) 171 (arXiv: hep-th/0503236); Phys.
Lett. B550 (2002) 121 (arXiv: hep-th/0209140); J.Wosiek Int.J. Mod. Phys. A
20 (2005) 4484 (arXiv: hep-th/0410066); J.Wosiek, ''Vacua of Supersymmetric
Yang-Mills Quantum Mechanics''(arXiv: hep-th/0510025 ); ''Supersymmetric
Yang-Mills Quantum Mechanics'' (arXiv: hep-th/0204243);

\bibitem{15}  Chung-Wen Kao, Gouranga C. Nayak and W. Greiner Phys. Rev. D66
(2002) 034017 (arXiv: hep-ph/0102153); Dennis D. Dietrich, Gouranga C. Nayak
and W. Greiner Phys. Rev. D64 (2001) 074006 (arXiv: hep-ph/0007139); Qun
Weng, Chung-Wen Kao, Gouranga C. Nayak, H. Stoecker and W. Greiner Int. J.
Mod. Phys. E10 (2001) 483 (arXiv:hep-ph/0009076)

\bibitem{16}  R.G. Leigh, D. Minic and A. Yelnikov, ''Solving Pure
Yang-Mills in 2+1 Dimensions'' (arXiv: hep-th/0512111, appear in Phys. Rev.
Lett); ''How to Solve Pure Yang-Mills Theory in 2+1 Dimensions'' (arXiv:
hep-th/0604060)

\bibitem{17}  P. Gaete, E. Guendelman and E. Spalucci ''Confinement from
constant field condensates'' (arXiv: hep-th/0607151); P. Gaete and E.
Spalucci J. Phys. A39, 6021 (2006) (arXiv: hep-th/0512178); W.I. Weisberger
, Nucl. Phys. B161 (1979) 61; R. Parthasarathy, M. Singer and C. Viswanatham
, Canad. J. Phys. 61 (1983) 1442; T.N. Tudron, Phys. Rev. D22 (1980) 2566

\bibitem{18}  D. Ebert, V.V. Khudyakov, V.C. Zhukovsky and K.G. Klimenko,
Phys.\ Rev. D65, 054024 (2002) (arXiv: hep-ph/0106110); D. Ebert, V.C.
Zhukovsky and O.V. Tarasov, Phys.\ Rev. D72, 096007 (2005) (arXiv:
hep-ph/0507125); V. Skalozub and M.Bordag, Nucl. Phys. B576 (2000) 430
(arXiv: hep-ph/9905302); M. Bordag and V. Skalozub, Eur. Phys. J. C 45
(2006) 159 (arXiv:hep-th/0507141)

\bibitem{19}  L. Gendenshtein and I. Krive, Sov. Phys. Usp. 28 (1985) 645

\bibitem{20}  E. Witten, Nucl. Phys. B188 (1981) 513

\bibitem{21}  L. Gendenshtein, Sov. J. Nucl. Phys. 41 (1985) 166; JETP Lett.
39 (1984) 280

\bibitem{22}  F.Cooper, A. Khare and U. Sukhatme, Phys. Rept. 251 (1995) 267
(arXiv: hep-th/9405029)

\bibitem{23}  P. Salomonson and I. W. van Holten, Nucl. Phys. B196 (1982) 509

\bibitem{24}  A. Khare and J. Maharana, Nucl.Phys. B244 (1984) 409

\bibitem{25}  Sh. A. Mamedov, Candidate thesis, Moscow State University, 1991

\bibitem{26}  T. Banks, W. Fischler, S. Shanker and L. Susskind, Phys. Rev.
D55 (1997) 6189 (arXiv: hep-th/9610043)

\bibitem{27}  V.Ch. Zhukovskii and Yu. N. Belousov, Sov. Phys. J. 32 (1989)
109

\bibitem{28}  R. Flume, Ann. Phys. 164 (1985) 189

\bibitem{29}  Sh. Mamedov, ''Energy States of Colored Particle in a
Chromomagnetic Field'', (arXiv: hep-th/0608142, v.1); ''Diagonalization of the Hamiltonian in a Chromomagnetic Field'' (arXiv: hep-th/0611267)
\end{thebibliography}
\end{document}